\documentclass[preprint,12pt,authoryear]{elsarticle}

\let\today\relax
\makeatletter
\def\ps@pprintTitle{%
    \let\@oddhead\@empty
    \let\@evenhead\@empty
    \def\@oddfoot{\footnotesize\itshape
         {Accepted Manuscript in European Journal of Mechanics - A/Solids} \hfill\today}%
    \let\@evenfoot\@oddfoot
    }
\makeatother

\usepackage{amssymb}

\usepackage{lineno}
\usepackage[hidelinks]{hyperref}
\hypersetup{colorlinks,bookmarksopen,linkcolor=blue,citecolor=blue,urlcolor=blue}
\usepackage{xcolor}
\usepackage{fullpage}

\journal{Elsevier}

\usepackage[english]{babel}
\usepackage[utf8x]{inputenc}
\usepackage[T1]{fontenc}
\usepackage{bm}

\usepackage{pgfplotstable}
\usepackage{pgfplots}
\usepackage{tikz}
\usepackage{stanli}
\usepackage{amsmath}
\usepackage{amsfonts}
\usepackage{url}
\pgfplotsset{compat = 1.15}
\usepackage{graphicx}
\usepackage{color}
\usepackage{geometry}
\usepackage{setspace}
\usepackage{siunitx}
\usepackage{multirow}
\usepackage{graphicx}
\usepackage{changepage}
\usepackage{chngcntr}

\usetikzlibrary{external}
\usetikzlibrary{angles,quotes}
\usetikzlibrary{arrows.meta}
\usepackage{soul} 
\soulregister\ref{7}
\soulregister\citep{7}
\soulregister\cite{7} 

\begin{document}

\begin{frontmatter}



\title{Towards active stiffness control in pattern-forming pneumatic metamaterials}%


\author[ctumech,tuemom]{Ondřej Faltus\corref{cor1}}
\ead{ondrej.faltus@cvut.cz}
\cortext[cor1]{Corresponding author}
\author[ctumech]{Milan Jirásek}
\author[ctumech,casita]{Martin Horák}
\author[ctumech]{Martin Doškář}
\author[tuemom]{Ron Peerlings}
\author[ctumech]{Jan Zeman}
\author[tuemom]{Ondřej Rokoš}

\affiliation[ctumech]{
            organization={Department of Mechanics, Faculty of Civil Engineering, Czech Technical University in Prague},
            addressline={Thákurova 7}, 
            postcode={166 29},
            city={Prague 6},
            country={Czech Republic}}

\affiliation[tuemom]{
            organization={Mechanics of Materials, Department of Mechanical Engineering, Eindhoven University of Technology},
            addressline={P.O. Box 513}, 
            postcode={5600MB},
            city={Eindhoven},
            country={The Netherlands}}

\affiliation[casita]{
            organization={Institute of Information Theory and Automation, Czech Academy of Sciences},
            addressline={Pod Vodárenskou věží 4}, 
            postcode={182 00},
            city={Prague 8},
            country={Czech Republic}}

\begin{abstract}
Pattern-forming metamaterials feature microstructures specifically designed to change the material's macroscopic properties due to internal instabilities. These can be triggered either by mechanical deformation or, in the case of active materials, by other external stimuli, such as pneumatic actuation. We study a two-dimensional rectangular lattice microstructure which is pneumatically actuated by non-uniform pressure patterns in its voids, and demonstrate that this actuation may lead to different instability patterns. The patterns are associated with a significant reduction in the macroscopic stiffness of the material. The magnitude of this reduction can be controlled by different arrangements of the pressure actuation, thus choosing the precise buckled shape of the microstructure. We develop an analytical model and complement it with computational tests on a two-dimensional plane-strain finite element model. We explain the phenomenon and discuss ways of further developing the concept to actively control the stiffness of materials and structures.
\end{abstract}



\begin{keyword}
Mechanical metamaterials \sep Pneumatic actuation \sep Active control \sep Pattern-forming materials \sep Stiffness control
\end{keyword}

\end{frontmatter}


\section{Introduction}


Metamaterials are man-made materials engineered to exhibit properties that go beyond those of naturally occurring materials. While originally intended for applications in electromagnetics to achieve unnatural refraction or wave transmission properties~\citep{Ren2018, smith2004metamaterialsscience, Zheludev2010roadahead}, the concept nowadays extends to optics \citep{Ramakrishna2005emmetamatreview}, acoustics~\citep{Krushynska2014locallyresonantacoustics,Moleron2015acousticedgedetection,Tian2019acousticmetasurfaces}, or mechanics~\citep{Lee2012micronanometamatreview}. Metamaterials that are able to switch their unusual properties during their lifetime are known as active metamaterials or metadevices \citep{Xiao2020activereview}. To achieve this control, some form of actuation needs to be used. Among the most common are electric \citep{chen2017hybrid, yi2019piezoelectric}, magnetic~\citep{montgomery2021magneto}, pneumatic \citep{Matia2023pressureinsoftroboticactuators, Yang2015Bertoldigripper}, and internal stress~\citep{liu2019tensegrity} methods of actuation. Recent design advances also promote automated or at least semi-automated design of metamaterial microstructures, exploiting tessellation algorithms \citep{Goswami2019}, neural networks \citep{Ma2020}, or topology optimization \citep{hammer2000topoptusingforces, Tyburec2022modularTO}, leading possibly to modular manufacturing \citep{Doskar2023modularL}.

As a subgroup, mechanical metamaterials are designed to bring about desired mechanical properties at the macroscale. Negative Poisson's ratio, commonly known as auxeticity, was one of the first exotic behaviors to be achieved in this fashion \citep{Lakes1987negativepoisson, Ren2018}. It can be induced by, e.g., chirality of the microstructure~\citep{Alderson2010femforchiralhoneycombs, virk2013silicomb} or by pattern-forming behavior \citep{Bertoldi2010auxeticityinpatterning}. Apart from auxeticity, a large range of metamaterials exhibits variable macroscopic stiffness \citep{yu2018metamatreview}. Origami and kirigami microstructures, modeled after Japanese paper folding art, exhibit negative stiffness \citep{virk2013silicomb}, great stiffness reduction \citep{hwang2018tunable}, or a selection between loading paths for a stackable 3D unit cell, leading to different stiffnesses~\citep{Zhai2018origamitwopaths}. {Variable stiffness regions in a finite metamaterial 3D sample lead to effective vibration control~\citep{Zolfagharian2023carseatspring}, with global buckling modes suppressed by introducing the buckling sequentially}~\citep{Zolfagharian2022reentrantspring}{. A similar treatment of torsional buckling has been also recently addressed in }\cite{Ghorbani2024supressbucklingintorsion}{.} Tensegrity structures rely on internal stress to tune mechanical as well as acoustic properties \citep{liu2019tensegrity}. In addition, the bending stiffness of beam-like metamaterials can be manipulated by electric actuators \citep{chen2017hybrid, yi2019piezoelectric}. Some metamaterials also use a surrounding magnetic field to change the effective mechanical properties of the microstructure \citep{montgomery2021magneto}.

Pneumatically actuated metamaterials rely on voids or gaps in the microstructure, which are pressurized or depressurized to change the effective metamaterial properties. This can be utilized in sandwich designs for electromagnetic applications \citep{khodasevych2012reconfigurablefishnet, su2020dualbandpneumaticabsorber}, noise absorbers with tunable acoustic band gaps \citep{Hedayati2020acousticpneumaticactuated}, manufacturing of optical metamaterials \citep{Feng2018opticalpressureactuatedchiral}, or pneumatically activated gripper mechanisms \citep{Yang2015Bertoldigripper}. Mechanical metamaterials can also be pneumatically actuated to exhibit negative stiffness and act as vibration absorbers \citep{tan2020realtimenegstiff} or bend upon command in a predetermined direction \citep{pan2020bendingactuator}.


Pattern-forming mechanical metamaterials are typically two-dimensional metamaterials whose microstructure consists of periodically arranged holes in an elastomeric sheet or a slab. Upon mechanical loading they undergo pattern transformation, i.e. localized buckling of the ligaments between the voids. This behavior places them into the class of metamaterials that exploit microstructural instabilities \citep{Kochmann2017}. The instability is to be promoted here, rather than avoided, as it leads to interesting behavior, such as auxetic~\citep{Bertoldi2010auxeticityinpatterning} or even programmable \citep{Florijn2014} responses. Materials in this class may be distinguished mainly by the shape and arrangement of the voids in the two-dimensional sheet. Among the most prominent examples are (i) square-stacked circular holes \citep{Mullin2007patterning}, forming a simple auxetic material with a single pattern, and (ii) a hexagonal, honeycomb arrangement of circular holes, which features multiple patterns and buckling modes, leading to variable macroscopic properties \citep{papka1999honeycombcrushing, papka1999honeycombcrushinganalysis}. 
Notable applications are a variable Poisson's ratio material achieved by stacking multiple honeycomb microstructures \citep{Francesconi2019} and a simple soft robotic gripper~\citep{Yang2015Bertoldigripper}.

\begin{figure}
    \centering
    \includegraphics[width=.9\linewidth]{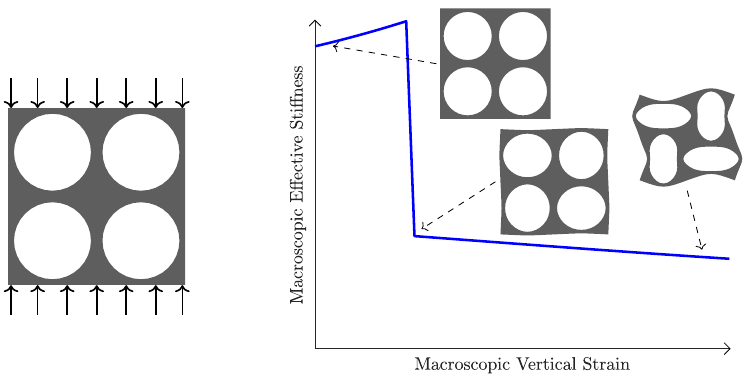}
    \caption{An illustration of a pattern-forming metamaterial with circular voids. Compressive macroscopic strain leads to an abrupt development of a local pattern associated with a drop in effective macroscopic stiffness.}
    \label{fig:patterns_on_circular_voids}
\end{figure}

Consider the square stacking of circular voids in the microstructure in Figure~\ref{fig:patterns_on_circular_voids}. As a result of macroscopic compressive strain the RVE develops a local pattern which is associated with a dramatic drop in the effective stiffness of the material. This behavior is elastic and fully reversible \citep{Mullin2007patterning}.

\begin{figure}[t]
    \centering
    \includegraphics[width=.8\linewidth]{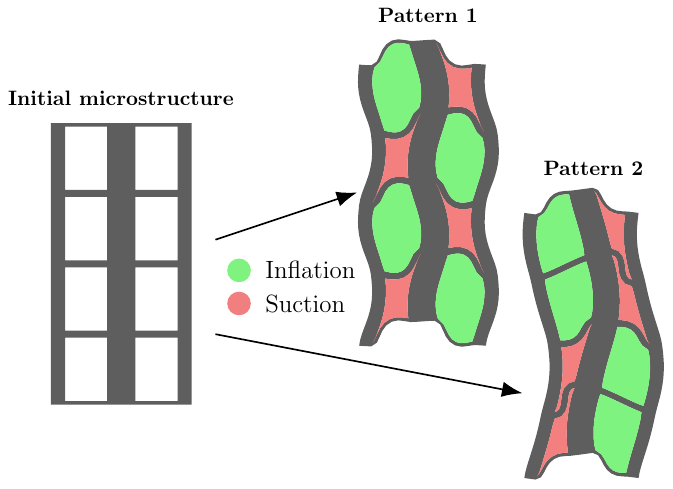}
    \caption{The periodicity of local patterns developed on a square lattice microstructure in response to pneumatic actuation corresponds to the periodicity of the applied pressurizing scheme. The same microstructure (undeformed shape, left) buckles into different patterns when the voids are pressurized in different configurations of positive (colored green) and negative (colored red) pressure.}
    \label{fig:motivation_RVE_and_patterns}
\end{figure}

In this contribution we focus on pneumatic pressure as a means of actuation for 2D pattern-forming metamaterials; a viable alternative to applied macroscopic strain~\citep{Chen2018pneumaticpatterns, Hyatt2022rapidpneumaticcontrol, Verhoeven2022}. The main motivation is that triggering a pattern solely by pneumatic actuation would mean retention of the associated changes in stiffness and auxeticity without a need for an externally prescribed macroscopic strain field. Moreover, if a material could be chosen with multiple actuatable patterns and fully elastic reversible behavior, an active choice of the altered mechanical parameters would be made possible. In this paper we consider a square lattice microstructure consisting of a periodical arrangement of rectangular voids in a polymer sheet. As to our knowledge and experience, materials with {equally sized} rectangular voids {arranged in a square lattice, i.e., with prismatic ligaments,} are typically prone to exhibit global buckling under macroscopic strain loading{, unlike their counterparts with circular voids~\citep{Oudes2021masterthesis}}. However, we demonstrate that pneumatic actuation can successfully induce patterning behavior. Figure \ref{fig:motivation_RVE_and_patterns} shows a representative volume element (RVE) of $2\times4$ rectangular voids and the way it forms different patterns based on the nonuniform schemes of air pressure applied to the voids (depicted by different colors). The microstructure represented by this RVE could also be considered as a series of horizontally repeating vertical columns, onto which a horizontal deflection is imposed. Intuition would suggest, and this contribution will later confirm, that their stiffness in the vertical direction would then diminish with the second power of the horizontal deflection amplitude. Since this amplitude is different between the two depicted pressurization schemes for the same reference sample, this actuation promises a mechanism to actively control the vertical macroscopic stiffness of the metamaterial. It should be mentioned that this reasoning does not yet take into account the presence of the horizontal ligaments connecting the columns, which turns out to limit the stiffness reduction, as discussed later in this work. {We have chosen to focus on this particular geometry for its ease of buckling into the chosen pattern (see Figure~\ref{fig:motivation_RVE_and_patterns}). Furthermore, this geometry is in fact a lattice of prismatic beams, which is convenient for our analytical description of the buckling process presented in Section~\ref{sec:analytics}.}

This paper examines this phenomenon in several steps. Firstly in Section~\ref{sec:analytics}, we develop an analytical model based on beam theory to gain understanding of the mechanism of the instability and its dependence on the pneumatic pressure. Then in Section~\ref{sec:model}, we propose a metamaterial design based on these findings, which exhibits programmable changes in macroscopic stiffness, leading in Section~\ref{sec:simulations} to a computational demonstration of switchable stiffness behavior on a unit cell comprising $2 \times 8$ rectangular voids. Finally, in Section~\ref{sec:conclusions}, we collect concluding remarks and a discussion of possible further development of the concept.

\section{Analytical models for pneumatically actuated rectangular lattices}
\label{sec:analytics}

\begin{figure}
    \centering
    \resizebox{\linewidth}{!}{
    \begin{tabular}{c c c}
         \includegraphics[width=.4\linewidth]{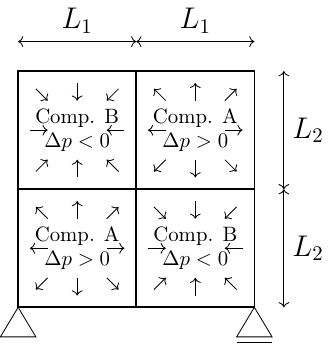}
         &
         \includegraphics[width=.4\linewidth]{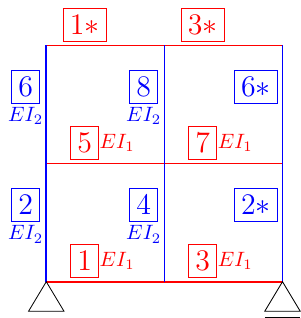}
         &
         \includegraphics[width=.4\linewidth]{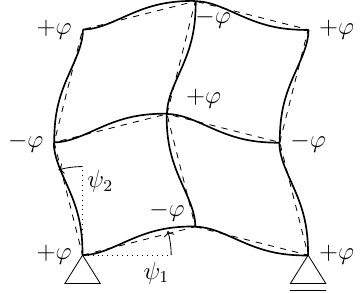}
         \\
         (a) & (b) & (c)
    \end{tabular}
    }
    \caption{An analytical description based on beam theory: (a) the initial geometry of four compartments labeled A and B in a checkerboard pattern, their pneumatic loading, and boundary conditions, (b) numbering of beams and their flexural characteristics, with periodic images of the same beams denoted by asterisks, (c) deformed geometry (dashed lines correspond to beam chords), definition of chord inclination angles $\psi_1$ and $\psi_2$ and joint rotation angles $\varphi$ and $-\varphi$.}
    \label{fig:analytical_lattice}
\end{figure}

A two-dimensional periodic unit cell of $2\times2$ rectangular voids can be modeled as consisting of beams, as depicted in Figure~\ref{fig:analytical_lattice}a. Here we construct an analytical approximation based on a geometrically nonlinear beam model with moderate rotations to examine the basic features of the response of such a microstructure to pneumatic actuation, including the emergence of an internal instability that causes the development of an internal pattern directly related to the applied pneumatic loading. The following analysis leads to sufficiently accurate estimates only for square lattices composed of prismatic and slender beams. Nevertheless, it allows for the identification of important parameters governing the patterning process and exploring their scaling. The analysis also provides insight into and permits a qualitative description of the basic trends that can be expected for microstructures with thicker ligaments.

In the following, we lay out the assumptions reducing the periodic unit cell model to only three degrees of freedom (Section \ref{sec:analytic_problem_setting}), and then examine the tangential stiffness of the system to determine the critical value of pneumatic loading at which internal stability is lost (Section~\ref{sec:analytic_crit_p}). In Section~\ref{sec:physical_meaning}, we offer an explanation as to the physical origin of the observed behavior. The results are verified against numerical models in Section~\ref{sec:critical_pressure_comparison}, and limitations of the beam theory are discussed. Finally, in Sections~\ref{sec:pressure_and_vertical_stress} we extend the analytical model such that it accounts for loading by vertical macroscopic strain, with an example calculation provided then in Section~\ref{sec:pressure_and_vertical_stress_example}, and in Section \ref{sec:csl} we predict a critical state curve characterizing the effect of an interplay of pneumatic and macroscopic strain loading on the development of the instability.

\subsection{Problem setting}
\label{sec:analytic_problem_setting}

Consider a regular rectangular lattice of flexible beams connected by rigid joints as depicted in Figure \ref{fig:analytical_lattice}a. Horizontal beams are characterized by their length $L_1$ and flexural sectional stiffness $EI_1$, while vertical beams are of length $L_2$ and stiffness $EI_2$ (see Figure~\ref{fig:analytical_lattice}b), with $E$ representing the Young's modulus of the linear elastic material and $I$ the in-plane moment of inertia of the beam. All beams are considered as axially inextensible and their shear distortion is neglected. The out-of-plane thickness, $t$, will be taken into account only for the sake of dimensional consistency. The initially rectangular void compartments bounded by the beams are subjected to prescribed pressure differences $\Delta p$ (inflation) and $-\Delta p$ (suction) alternating in a checkerboard pattern. Over-pressurized and under-pressurized compartments are referred to as A and B, respectively. Based on the assumed periodicity, it is expected that the deformed shapes of all A compartments are identical, except for an opposite sign of rigid rotation in neighboring layers, see Figure \ref{fig:analytical_lattice}c, and that the same holds for all B compartments. The inclination of all beam chords can thus be described by two angles: $\psi_1$ and $\psi_2$. Horizontal beam chords rotate by an angle~$\psi_1$, counterclockwise in the left two compartments of the cell and clockwise in the right two compartments. Vertical beam chords rotate by an angle~$\psi_2$, again in the positive or negative sense depending on their position in the bottom or top pair of compartments. By symmetry, all of the rigid joints in the unit cell might rotate by angles of the same magnitude $\varphi$, with the sense of the rotation alternating in a checkerboard pattern as shown in Figure \ref{fig:analytical_lattice}c. In this analysis, counterclockwise rotations are considered to be positive.

Based on the assumption of beam axial incompressibility, all joint displacements can be calculated from the rotations $\psi_1$, $\psi_2$, and $\varphi$. These three variables thus fully describe the state of the microstructure, and the stability of the resulting model with three degrees of freedom can be evaluated analytically, taking into account the influence of the prescribed pressure difference $\Delta p$. 

\subsection{Calculation of the critical pressure difference}
\label{sec:analytic_crit_p}

For all vertical beams, the rotation of their ends with respect to the chord is either $\varphi-\psi_2$ at one end and $-\varphi-\psi_2$ at the opposite end, or $-\varphi+\psi_2$ at one end and $\varphi+\psi_2$ at the opposite end, which means that their deformed shapes are equivalent and the contribution 
of each of them to the strain energy is the same. A similar argument is valid for all the horizontal beams.
As derived in detail in \ref{app:unit_cell_analytical_model}, the total strain energy $\mathcal{E}_\mathrm{int}$ of the periodic cell is then
\begin{equation}\label{mj1}
    \mathcal{E}_\mathrm{int}(\varphi,\psi_1,\psi_2)=  8k_1 (\varphi^2+3\psi_1^2) + 8k_2 (\varphi^2+3\psi_2^2)
\end{equation}
 where $k_1 = EI_1/L_1$ and $k_2 = EI_2/L_2$ are the flexural beam stiffnesses.

External forces consist of the prescribed pressure difference $\Delta p$, which is work-conjugate to the compartment volume changes $\Delta V_A$ and $\Delta V_B$. The latter can also be expressed in terms of the three degrees of freedom, leading to the  energy of external forces given by
\begin{equation}\label{eq:mj2}
    \mathcal{E}_\mathrm{ext}(\varphi,\psi_1,\psi_2) = \frac{4}{3}(L_1^2-L_2^2)t\,\Delta p\,\varphi-
     4\,L_1^*L_2^* t\,\Delta p \sin\psi_1\sin\psi_2
\end{equation}
 where $L_1^* = L_1 + \Delta L_1^*$ and $L_2^* = L_2 + \Delta L_2^*$ are the beam chord lengths (i.e., distances between the end joints) in the deformed state. As explained in \ref{app:unit_cell_analytical_model}, the dependence of the increments $\Delta L_1^*$ and $\Delta L_2^*$ on the three degrees of freedom and the applied pressure difference can be approximated by quadratic functions specified in (\ref{app24})--(\ref{app25}).

Setting the partial derivatives of the total potential energy $\mathcal{E}_\mathrm{p} = \mathcal{E}_\mathrm{int}+\mathcal{E}_\mathrm{ext}$ equal to zero, we obtain the following set of three nonlinear equilibrium equations
\begin{eqnarray}\nonumber 
    16(k_1+k_2)\varphi+\frac{4}{3}(L_1^2-L_2^2)t\,\Delta p + \frac{4}{3} (L_1L_2^*+L_1^*L_2)t\,\Delta p\,\varphi\sin\psi_1\sin\psi_2
    +\\ \label{mj3}
    +\frac{1}{45}\left(\frac{L_1^*L_2^3}{k_2}-\frac{L_1^3L_2^*}{k_1}\right)\sin\psi_1\sin\psi_2
    &=&0\;\;\;\; \\ \label{mj4}
    48k_1\psi_1 -4L_1^*L_2^*t\,\Delta p \cos\psi_1\sin\psi_2 +\frac{4}{5}L_1L_2^*t\,\Delta p\,\psi_1\sin\psi_1\sin\psi_2 &=& 0\;\;\;\;
    \\ \label{mj5}
 48k_2\psi_2 -4L_1^*L_2^*t\,\Delta p \sin\psi_1\cos\psi_2 +\frac{4}{5}L_1^*L_2t\,\Delta p\,\psi_2\sin\psi_1\sin\psi_2 &=& 0\;\;\;\;
\end{eqnarray}
These equations admit a fundamental solution characterized by $\psi_1=\psi_2=0$ and
\begin{equation}\label{eq:firstsimplesol}
\varphi = \frac{L_2^2-L_1^2}{12(k_1+k_2)}t\,\Delta p
\end{equation}
 which corresponds to a deformation pattern with all beam chords remaining either horizontal or vertical, and the joint rotation growing proportionally to the applied pressure difference.

The fundamental solution exists for arbitrary $\Delta p$, but its stability can be lost at a certain pressure level. This stability can be assessed based on the tangent stiffness, i.e., the Hessian of the total potential energy $\mathcal{E}_\mathrm{p}$, represented in this case by a matrix of second-order derivatives. For the fundamental solution above, the tangent stiffness matrix reads
\begin{equation}
    \mathsf{K} = \left(\begin{array}{ccc}
        16k_1+16k_2  &  0 & 0 \\
        0  & 48k_1 &  -4L_1^*L_2^*t\,\Delta p \\
        0 & -4L_1^*L_2^*t\,\Delta p & 48k_2
    \end{array}\right)
\label{eq:tangent_pressure}
\end{equation}
This matrix remains positive definite as long as the determinant of the lower right $2\times 2$ submatrix is positive, i.e., as long as
\begin{equation}
    (48k_1)(48k_2)-(4L_1^*L_2^*t\,\Delta p)^2>0
\end{equation}
The above condition is violated for the first time when the pressure difference attains its critical value
\begin{equation}\label{eq:dpcrit}
    \Delta p_\mathrm{crit}^{(0)}=\frac{12\sqrt{k_1k_2}}{L_1^*L_2^*t}
\end{equation}
The superscript $\bullet^{(0)}$ refers to the fact that this estimate of critical pressure difference
has been obtained at zero externally applied stress.

Recall that the terms $L_1^*$ and $L_2^*$ themselves depend on $\Delta p$, and so the precise determination of the critical pressure difference requires an iterative approach. However, if the loss of stability occurs relatively early, one can start from an initial estimate $L_1^*=L_1$ and $L_2^*=L_2$ and then proceed by substituting Equations~(\ref{app24}) and (\ref{app25}) into Equation (\ref{eq:dpcrit}). The resulting nonlinear equation can be solved numerically for the exact values of $\Delta  L_1^*$ and $\Delta L_2^*$.

For illustration,  consider the most regular case when all beams have the same properties, i.e., $L_1=L_2=L$, $EI_1=EI_2=EI$ and thus $k_1 = k_2 = k$. As long as $\psi_1=\psi_2=0$ (i.e., for the fundamental solution), we have $\varphi=0$ following Equation~(\ref{eq:firstsimplesol}), and Equations (\ref{app24})--(\ref{app25}) are then simplified with the change in chord length given by
\begin{equation}\label{eq10}
  \Delta  L^* =- \frac{L^5}{15120}\left(\frac{t\,\Delta p}{k}\right)^2 
\end{equation}
In this particular case, the critical state given by Equation (\ref{eq:dpcrit}) is reached at the pressure difference of
\begin{equation}\label{eq:dpcrit0}
    \Delta p_\mathrm{crit}^{(0)}=\frac{12k}{t(L^*)^2}
\end{equation}
\noindent Substituting this expression into Equation (\ref{eq10}), we obtain an implicit formula for evaluating the change in chord length at the critical state,
\begin{equation}\label{eq:dpcrit0x}
  \Delta  L^* =- \frac{L^5}{15120}
  \left(\frac{12}{(L+\Delta L^*)^2}\right)^2 =  
  - \frac{L^5}{105(L+\Delta L^*)^4}
\end{equation}
\noindent Solving this nonlinear equation yields $\Delta L^*=-9.9109\times 10^{-3}L$.

Substituting $L^*=L+\Delta L^*=0.9901\,L$ into Equation~(\ref{eq:dpcrit0}) finally yields the resulting improved estimate of the critical pressure difference $\Delta p_\mathrm{crit}^{(0)}=12k/(t(0.9901\,L)^2) =12.241\, k/tL^2$. Note that $\Delta L^*/L$ represents the macroscopic normal strain at the onset of instability, which is in the present case below $1~\%$ (in horizontal as well as vertical direction). This strain represents the contraction of the microstructure caused by pressurization up to the stability limit.

 
When the fundamental solution loses stability, bifurcation into a different mode can be expected. The direction of the increment at the onset of bifurcation is determined by the eigenvector that corresponds to the zero eigenvalue of the tangent stiffness matrix. In the particular case of square geometry with identical properties of all beams, this eigenvector is $(\varphi,\psi_1,\psi_2)=(0,1,1)$. This means that we can expect a sudden appearance of chord rotations with equal magnitudes for $\psi_1$ and $\psi_2$, as seen in the schematic plot in Fig.~\ref{fig:analytical_lattice}c. 

\subsubsection{Physical origin of the instability}
\label{sec:physical_meaning}

\begin{figure}[t]
    \centering
    \resizebox{0.9\linewidth}{!}{
    \begin{tabular}{c c}
         \includegraphics[width=.5\linewidth]{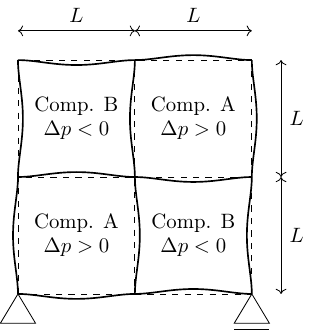}
         &
         \includegraphics[width=.5\linewidth]{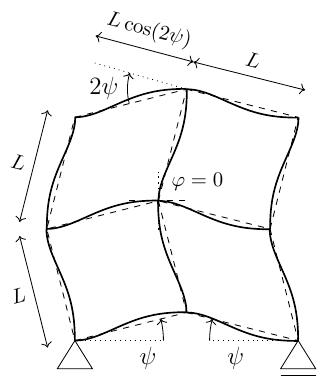}
         \\
         (a) & (b)
    \end{tabular}
    }
    \caption{Illustration of an analytical solution to a simplified problem with $L_1=L_2=L$ and $EI_1=EI_2=EI$ (a) The pre-bifurcation state with comparments labeled and loading indicated (deflection magnified by a factor of 1.5). For the fundamental solution the chord (dashed lines) rotations $\psi_1=\psi_2=0$, and in this particular case, $L_1 = L_2$ leads to the joint rotations $\varphi=0$ as well, recall Equation (\ref{eq:firstsimplesol}) and Figure~\ref{fig:analytical_lattice}c. (b) The bifurcated stable solution, where all beam chords rotate by $\psi$ as pictured, but $\varphi = 0$ still.}
    \label{fig:simple_solution}
\end{figure}

When we focus on the essential terms in the above derivation, the physical origin of the predicted instability can be unraveled. In the simplest case of a square equi-axial geometry with $L_1=L_2=L$ and $EI_1=EI_2=EI$, neither the chords nor the joints (recall Equation (\ref{eq:firstsimplesol})) rotate before the fundamental solution loses stability. Therefore, if we plot only the chords, the basic pattern remains composed of squares, as illustrated in Figure \ref{fig:simple_solution}a. The bifurcated solution involves rotations of the chords such that the chord outlines of the A compartments (inflated) keep their square shape and only rotate while the chord outlines of the B compartments (under suction) fold into diamond shapes, see Figure~\ref{fig:simple_solution}b. The driving force behind the instability is thus the reduced area of the B compartments; recall that the external energy involves the work of the applied pressure difference on the volume change, see Equation (\ref{eq:mj2}). If the chords rotate by $\psi$ as indicated in Figure~\ref{fig:simple_solution}b, the volume of the diamond-shaped B compartment can be expressed as
\begin{equation}\label{mj13}
    V_\mathrm{B} = tL^2\cos 2\psi
\end{equation}

\noindent where $t$ is the out-of-plane thickness. The first derivative of $V_\mathrm{B}$ with respect to $\psi$ vanishes, but the second derivative is negative, which can give a negative contribution to the second-order derivative of the external work if the compartment is under negative pressure difference (suction). Note that the A compartments do not change their volume at all.

Since the beams have rigid connections at the joints, they resist the folding process by their bending stiffness. Each of these beams deforms in the same way as if one end is fixed and the other experiences a lateral deflection $w=L\sin\psi$ while its rotation remains zero. Elementary beam analysis leads to the conclusion that the elastic energy stored in such a beam is given by $6EIw^2/L^3$. When a rotation $\psi$ occurs, we can express the increase in the potential energy of the entire periodic cell, based on the above considerations, as follows:
\begin{equation}
    \Delta \mathcal{E}_\mathrm{p} = 8 \cdot \frac{6EI(L\sin\psi)^2}{L^3} + 2t  L^2(\cos 2\psi-1)\Delta p
\end{equation}
\noindent Here, we have considered that the cell contains 8 full beams and that the cell contains two B compartments that fold, while the two A compartments do not change their volume. Also, we have replaced volume $V_\mathrm{B}$ by its increment $\Delta V_\mathrm{B} = V_\mathrm{B}-tL^2$ caused by the chord rotation $\psi$. It is essential that the contribution of applied pressure difference to the energy is $\Delta p \Delta V$, with a positive sign, because the pressure difference in the B compartments is negative, i.e., $-\Delta p$. Therefore, positive work is supplied by the applied suction if the compartment area is diminished.

To detect the critical pressure difference, one option is to take the second derivative of $\Delta \mathcal{E}_\mathrm{p}$ with respect to $\psi$ and set it to zero. To achieve further insight into the role of the variables involved, we can obtain the same result in an alternative way. We look at the leading terms in the Taylor expansion of $\Delta \mathcal{E}_\mathrm{p}$ at $\psi=0$ and replace $\sin\psi$ by $\psi$ and $\cos 2\psi-1$ by $-2\psi^2$. The resulting second-order approximation of the energy increment then reads
\begin{equation}\label{mj15}
    \Delta \mathcal{E}_\mathrm{p} \approx \frac{48EI\psi^2}{L} - 4t  L^2\psi^2\Delta p = \left(\frac{48EI}{L} - 4t  L^2\Delta p\right)\psi^2
\end{equation}

\noindent It is obvious from this expression that the bending stiffness represented by the factor $48EI/L$ has a stabilizing effect while the negative pressure contribution represented by the factor~$4t L^2\Delta p$ has a destabilizing effect. The overall factor multiplying $\psi^2$ ceases to be positive when the pressure difference attains its critical value, given by $12EI/(tL^3)$. Note that this is almost the same result as in Equation~(\ref{eq:dpcrit0}), except for the difference between $L$ and $L^*$. For simplicity here, we have used the same length $L$ for the evaluation of the beam elastic energy (the first term in Equation~(\ref{mj15})) and for the evaluation of the reduced volume in Equation~(\ref{mj13}), reflected by the second term in Equation~(\ref{mj15}). In a refined calculation, we could evaluate the diamond-shaped compartment volume using the current chord length $L^*$ instead of the original undeformed length $L$, and then we would get exactly the same critical pressure difference as in Equation~(\ref{eq:dpcrit0}).

This mechanical reasoning for the simplest setup provides physical insight into the key mechanisms affecting the bifurcation. Still, the general derivation found in Section~\ref{sec:analytic_crit_p} and in \ref{app:unit_cell_analytical_model} is needed to cover other geometries and stiffness ratios.

\subsubsection{Comparison with numerical simulations}
\label{sec:critical_pressure_comparison}

The simplified analytical prediction of the critical pressure difference was verified numerically. To this end, we simulated the response of the unit cell in the OOFEM open-source software \citep{Patzak2001OOFEMAES}, utilizing a novel formulation of a geometrically exact nonlinear beam element developed in \cite{JLMRH21nonlinbeam}, which neglects shear effects but takes into account axial extensibility. The effect of lateral pressure, taking into account the geometry changes, had been implemented for the purpose of the present study as an additional feature. Note that the solution presented in Section~\ref{sec:analytic_crit_p} is based on flexural effects only, neglecting the deformation of beam segments by axial tension/compression and by shear distortion. Therefore, quantitative agreement can be expected only for microstructures comprised of sufficiently slender ligaments, for which the influence of axial deformation is negligible.
 

\begin{figure}[h]
    \centering
    \begin{tabular}{ccc}
         \includegraphics[width=0.3\textwidth]{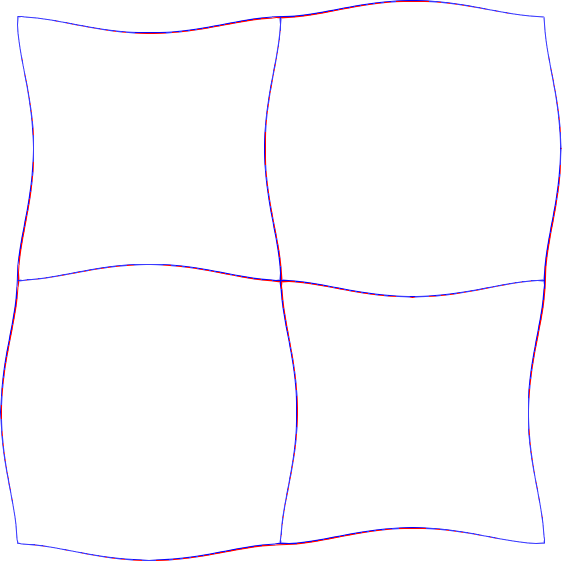}
         & 
         \includegraphics[width=0.3\textwidth]{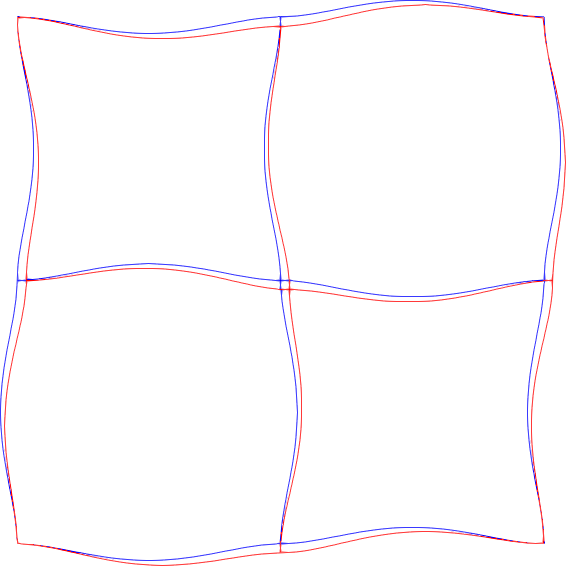}
         &
         \includegraphics[width=0.3\textwidth]{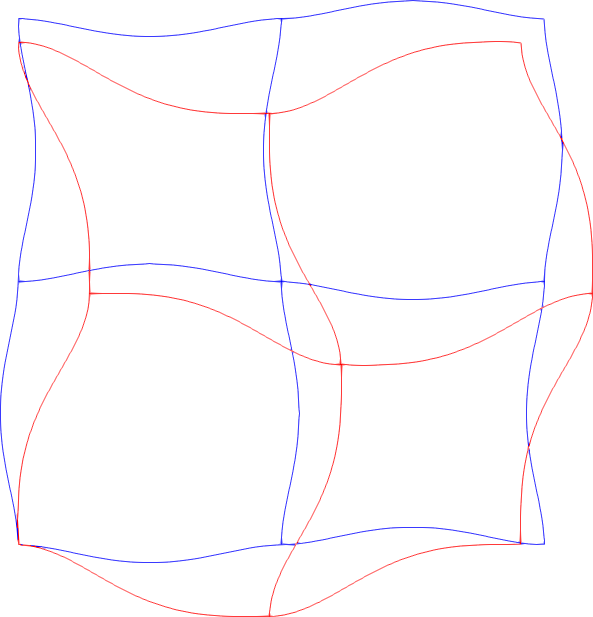}
         \\
         \footnotesize{(a)} & \footnotesize{(b)} & \footnotesize{(c)}
    \end{tabular}
    \caption{Results of a numerical simulation of a periodic cell loaded by alternating internal pressure difference: fundamental solution that becomes unstable (blue), and perturbed solution that closely follows a stable bifurcated branch (red) at pressure differences
    (a) \SI{2.25}{\kilo\pascal} (precritical), (b) \SI{2.30}{\kilo\pascal} (close to critical), and (c) \SI{2.50}{\kilo\pascal} (post-critical).}
    \label{f:cell2}
\end{figure}

The simulation shown here in Figure \ref{f:cell2} uses a square geometry of the cell, with $L_1=L_2=L=\SI{6}{\meter}$ and $t=\SI{1}{\meter}$. The flexural beam stiffnesses are set to $EI=\SI{40}{\newton\meter^2}$ and $EA=\SI{4000}{\newton}$, where $A$ is the beam cross section area and $EA$ thus represents the axial beam stiffness. This corresponds to the dimensionless parameter $EAL^2/EI=L^2/i^2=3600$ and slenderness ratio $L/i=60$, i.e., a reasonably slender beam (symbol $i$ denotes here the sectional radius of inertia, $i=\sqrt{I/A}$). The internal pressure is applied in the checkerboard pattern as shown in Figure~\ref{fig:analytical_lattice}a and its value is increased incrementally by $\SI{0.05}{\pascal}$ in each step. For this setup, the expected critical pressure difference evaluated from Equations (\ref{eq:dpcrit0}) and (\ref{eq:dpcrit0x}) is $\Delta p_\mathrm{crit}^{(0)} = {12.241\,EI/(tL^3) = \SI{2.267}{\pascal}}$, with the corresponding critical macroscopic strain $-0.0099$.

In the simulation, in order to detect the instability, analysis of the tangent stiffness matrix has been performed after each step. A negative eigenvalue of the stiffness matrix was first detected at pressure difference value of $\SI{2.35}{\pascal}$, when the minimum eigenvalue was -0.08855. In the previous step, i.e., at pressure difference $\SI{2.3}{\pascal}$, all eigenvalues were still positive and the smallest one was 0.00603. By linear interpolation we obtain an estimate of $\Delta p_\mathrm{crit}^{(0)}=2.303\;\si{\pascal}$, which differs from the analytical estimate, $\SI{2.267}{\pascal}$, only by 1.6~\%.

The simulation can be run up to very high pressure differences, exceeding the critical one. Without additional treatment, the iterative process converges, however, to the unstable solution, see the blue shape in Figure \ref{f:cell2}c. {This corresponds to the fundamental solution from Equation~(\ref{eq:firstsimplesol}) being followed beyond the critical point described by Equation~(\ref{eq:dpcrit}).} The stable, bifurcated solution can be followed numerically if a small perturbation is applied. It turned out to be sufficient to apply a constant vertical force of magnitude $2\cdot 10^{-4}\;\si{\newton}$ at the central node. Up to step 45 (pressure difference $\SI{2.25}{\pascal}$), the solutions of the original problem and the perturbed one remained visually indistinguishable; see Figure~\ref{f:cell2}a. The state after step 46 (pressure difference $\SI{2.30}{\pascal}$) is still stable, but it is very close to the critical one, and the perturbation pattern distinctly appears; see Figure~\ref{f:cell2}b. In subsequent steps, the deviation of the perturbed solution grows and the expected internal pattern develops; see Figure~\ref{f:cell2}c.

The finite element simulation on slender beams thus confirms the findings of the analytical stability model. The tangent stiffness loses positive definitiveness at a level of pressure loading closely correlated with the analytical prediction; the developed stable solution branch corresponds to the expected pattern.

\subsection{Interaction with macroscopic compression}
\label{sec:pressure_and_vertical_stress}

In the previous sections, we have solely considered loading by alternating inflation/suction applied in compartments A and B. Let us now add the effect of macroscopic normal nominal stress $\sigma_2$ applied in the vertical direction.
The added stress plays here the role of prescribed external loading, same as the pneumatic pressure difference $\Delta p$. Compressive stresses are considered, with $\sigma_2 < 0$. As shown in detail in \ref{app:pressure_stress_interaction}, the expression for the energy of external forces needs to be augmented with the term
\begin{equation}\label{mj16}
    \mathcal{E}_\mathrm{ext,\sigma} = -4L_1 \left(L_2^* \cos\psi_2 -L_2\right)t\sigma_2
\end{equation} 
\noindent Consequently, the left-hand sides of equilibrium equations (\ref{mj3}) and (\ref{mj5}) are correspondingly augmented by the partial derivatives of $\mathcal{E}_\mathrm{ext,\sigma}$, specified in (\ref{appb3})--(\ref{appb6}). It turns out that the fundamental solution characterized by $\psi_1=\psi_2=0$ still satisfies Equations (\ref{mj4})--(\ref{mj5}), even after the augmentation. However, the inclusion of the vertical loading affects the relation between $\varphi$ and $\Delta p$, determined from the modified version of Equation (\ref{mj3}). If we omit the terms that vanish for $\psi_1=\psi_2=0$ and replace $\cos\psi_2$ by 1, the bifurcation condition now reads
\begin{equation}
    16(k_1+k_2)\varphi+\frac{4}{3}(L_1^2-L_2^2)t\,\Delta p + \frac{4}{3}L_1t\sigma_2\left(L_2\varphi+\frac{L_2^3t\,\Delta p}{60k_2}\right) = 0
\label{mj3mod}
\end{equation}
\noindent and the resulting expression for the joint rotation is
\begin{equation}\label{mj18}
    \varphi = \frac{L_2^2-L_1^2-L_1L_2^3t\sigma_2/(60k_2) }{12(k_1+k_2)+L_1L_2t\sigma_2}\,t\,\Delta p
\end{equation}
\noindent This is a generalized version of Equation (\ref{eq:firstsimplesol}); the joint rotation is still proportional to the internal pressure, but the proportionality factor is now affected by the vertical macroscopic stress $\sigma_2$, and $\varphi$ does not vanish for $L_1 = L_2$.

Expression (\ref{mj18}) for the rotation $\varphi$ (along with $\psi_1=\psi_2=0$) can be substituted into Equation (\ref{app25}) to obtain the change in chord length of the vertical beams, from which the macroscopic vertical strain can be evaluated (taking into account again that $\psi_2=0$):
\begin{eqnarray}
    \varepsilon_2 = \frac{\Delta L_2^*}{L_2} &=& \left[ -\frac{1}{6}\left(\frac{L_2^2-L_1^2-L_1L_2^3t\sigma_2/(60k_2) }{12(k_1+k_2)+L_1L_2t\sigma_2}\right)^2 +\right. \\ \nonumber
    && +\left.\frac{L_2^2}{180k_2}\frac{L_1^2-L_2^2+L_1L_2^3t\sigma_2/(60k_2) }{12(k_1+k_2)+L_1L_2t\sigma_2}-\frac{L_2^4}{15120k_2^2}\right]t^2(\Delta p)^2
\end{eqnarray}
This is the inverted form of the stress-strain law describing the fundamental solution, which ignores the contribution of axial compressibility of the vertical beams.

The stability of the fundamental solution depends on the eigenvalues of the augmented tangent stiffness matrix, which is obtained by adding terms that correspond to the second-order derivatives of $\mathcal{E}_\mathrm{ext,\sigma}$ (elaborated in \ref{app:pressure_stress_interaction}) to the matrix in Equation (\ref{eq:tangent_pressure}). It turns out that only two diagonal entries are affected, see (\ref{appb10})--(\ref{appb11}). The resulting tangent stiffness matrix is given by
\begin{equation}
   \mathsf{K} = \left(\begin{array}{ccc}
   16k_1+16k_2+4L_1L_2t\sigma_2/3  &  0 & 0 \\
   0  & 48k_1 &  -4tL_1^*L_2^*\Delta p \\
   0 & -4tL_1^*L_2^*\Delta p & 48k_2 + 4L_1\left(L_2/5+L_2^*\right)t\sigma_2
\label{eq:tangent_pressureandstress}
\end{array}\right)
\end{equation}
\noindent For compression ($\sigma_2 < 0$), the diagonal stiffness coefficients $K_{11}$ and $K_{33}$ are reduced. For each of these coefficients its reduction leads to a separate and independent critical state condition. The first one is related to buckling of beams under compression, and we shall denote its critical value as $\sigma^{(b)}_\mathrm{2,crit}$; the second one is related to pneumatic pressure interaction with the critical value $\sigma^{(\Delta p)}_\mathrm{2,crit}$.

Coefficient $K_{11}$ ceases to be positive if $ \sigma_2 \le \sigma^{(b)}_\mathrm{2,crit}$ where
\begin{equation}
\label{eq:critstress}
     \sigma^{(b)}_\mathrm{2,crit} = -\frac{12(k_1+k_2)}{tL_1L_2}
\end{equation}
This potential critical state is independent of the pneumatic pressure and corresponds to standard buckling of beams under compressive loads.

The determinant of the lower $2\times 2$ submatrix in Equation~(\ref{eq:tangent_pressureandstress}) becomes non-positive for $\Delta p \ge \Delta p^{(\sigma)}_\mathrm{crit}$ where
\begin{equation} 
\label{eqpcrit2}
    \Delta p_\mathrm{crit}^{(\sigma)} = \frac{\sqrt{12k_1(12k_2+L_1\left(L_2/5+L_2^*\right)t\sigma_2)}}{tL_1^{*}L_2^{*}}
\end{equation}
\noindent This formula shows that the applied compressive stress reduces the critical pressure difference. Note that setting $\sigma_2=0$ recovers Equation (\ref{eq:dpcrit}). Alternatively, one could express the critical value of $\sigma_2$ as a function of applied $\Delta p$:
\begin{equation}\label{mj22}
    \sigma_\mathrm{2,crit}^{(\Delta p)} = \frac{(tL_1^*L_2^*\,\Delta p)^2-144k_1k_2}{12k_1tL_1\left(L_2/5+L_2^*\right)}
\end{equation}

For deeper insight, it is useful to write the interaction between the pneumatic pressure and the vertical macroscopic stress in a dimensionless format. In a simplified form, the critical state condition
%
can be rewritten as
\begin{equation}\label{eq:mj23}
   \frac{\sigma_\mathrm{2,crit}^{(\Delta p)}}{\sigma_\mathrm{2,crit}^{(0)}}+\left(\frac{\Delta p_\mathrm{crit}^{(\sigma)}}{\Delta p_\mathrm{crit}^{(0)}}\right)^2 = 1 
\end{equation}
where $\Delta p_\mathrm{crit}^{(0)}$ is the critical pressure difference at zero applied stress, defined in Equation~(\ref{eq:dpcrit}), and
\begin{equation}\label{eq:sigma_crit_0}
    \sigma_\mathrm{2,crit}^{(0)} = -\frac{60k_2}{tL_1(L_2+5L_2^*)}
\end{equation}
is the critical stress at zero internal pressure difference. For most reasonable geometries, this critical stress related to the lower submatrix in Equation~(\ref{eq:tangent_pressureandstress}) is lower in magnitude than the critical stress from Equation (\ref{eq:critstress}).
For example, in the case of $k_1 = k_2$, the $\sigma^{(b)}_\mathrm{2,crit}$ value would only become dominant after the change in chord length reaches $\Delta L_2^* = -0.7 L_2$, which is however way past the point where stability of the fundamental solution is lost.

It is worth noting that condition (\ref{eq:mj23})
is exactly equivalent with the original condition of zero
determinant only if the quantities in the numerators
are considered as dependent on the chord lengths $L_1^*$ and $L_2^*$
that correspond to the specific combination of pressure and stress
in the critical state. If the denominators are replaced by constants
evaluated respectively in the critical state at zero pressure difference
and in the critical state at zero stress, then condition (\ref{eq:mj23})
becomes an analytical quadratic approximation.  
This will be illustrated by the following example.

\subsubsection{Comparison with numerical simulations}
\label{sec:pressure_and_vertical_stress_example}

To check the accuracy of the derived interaction according to Equation (\ref{eq:mj23}), the numerical investigation of the critical state outlined in Section \ref{sec:critical_pressure_comparison} is extended to combinations of internal pressure and uniaxial compressive stress. In the simplest case of a square lattice, {we can substitute} the rough approximation $L_1^*=L_2^*=L_1=L_2=L$ {into Equations~(\ref{eq:dpcrit}) and (\ref{eq:sigma_crit_0}) to obtain} the estimated characteristic values of the critical quantities {in the form}
\begin{eqnarray}\label{mj26}
    \Delta p_\mathrm{crit}^{(0)} &\approx& \frac{12EI}{tL^3} \\
     \sigma_\mathrm{2,crit}^{(0)} &\approx& -\frac{10EI}{tL^3} = -\frac{5}{6}\Delta p_\mathrm{crit}^{(0)}\label{mj27}
\end{eqnarray}

\begin{figure}[t]
    \centering
    \includegraphics[width=.65\linewidth]{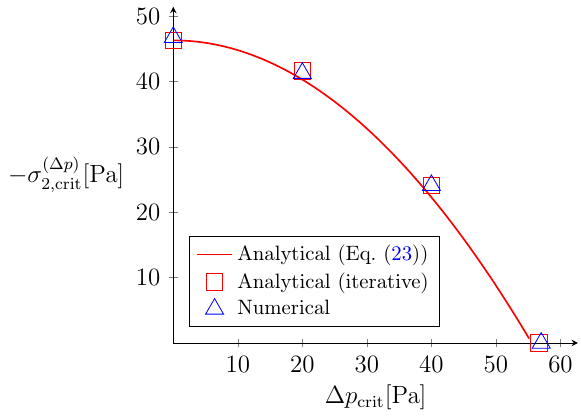}
    \caption{Combinations of pressure difference and macroscopic stress load leading to a bifurcation according to different models. A simplified analytical model with an approximation of chord length gives a quadratic curve in the $(\Delta p, \sigma_2)$ space, described by Equation (\ref{mj22}) and shown here in red. For selected loading combinations, improved estimates with an iterative estimation of chord lengths are depicted (red squares), demonstrating an even better correlation with numerical results from a beam simulation (blue triangles).}
    \label{fig:psig}
\end{figure}

Figure \ref{fig:psig} compares the presented analytical formulas with numerical results based on a beam model evaluated for the case when $t=\SI{1}{\meter}$, $h=\SI{1}{\meter}$, and $L=\SI{6}{\meter}$, with $EI= \SI{1 000}{\newton\meter^2}$ and $EA=\SI{12000}{\newton}$. The numerical results are represented as the blue triangles.


It is worth noting that for this case the critical stress evaluated from Equation (\ref{eq:critstress}) amounts to $|\sigma_\mathrm{2,crit}^{(b)}|= \SI{111.11}{\pascal}$ and is therefore much larger in magnitude than $|\sigma_\mathrm{2,crit}^{(0)}|$, which means that the stability limit is properly described by Equation (\ref{eq:mj23}), based on a vanishing determinant of the lower $2\times 2$ submatrix of the stiffness matrix. 

The simple analytical estimate is pictured as the red curve in Figure \ref{fig:psig}. The red squares represent its further improvement by evaluating $L_1^*$ and $L_2^*$ based on expressions (\ref{app24}) and (\ref{app25}). Substituting $L_1=L_2=L$ and $EI_1=EI_2=kL$ into~(\ref{mj18}), we obtain
\begin{equation}
    \varphi = -\frac{L^4t^2\sigma_2\,\Delta p }{60k(24k+tL^2\sigma_2)}
\end{equation}
\noindent which is then used in (\ref{app24})--(\ref{app25}), along with $\psi_1=\psi_2=0$. The resulting expressions for changes in the chord lengths as functions of the applied pressure difference and macroscopic stress read
\begin{eqnarray}
    \Delta L_1^*&=& 
    -t^2(\Delta p)^2\frac{L^5}{15120k^2}\left(1+\frac{1.4\,L^2t\sigma_2 }{24k+L^2t\sigma_2}+\frac{0.7\,L^4t^2\sigma_2^2 }{(24k+L^2t\sigma_2)^2} \right)\label{app24x}
\\
    \Delta L_2^* &=&  
  -t^2(\Delta p)^2\frac{L^5}{15120k^2}\left(1   -\frac{1.4\,L^2t\sigma_2}{24k+L^2t\sigma_2}+
   \frac{0.7\,L^4t^2\sigma_2^2 }{(24k+L^2t\sigma_2)^2} \right)\label{app25x}
\end{eqnarray}

\noindent It is now possible to evaluate the improved estimate iteratively. For given $\Delta p_\mathrm{crit}$, we first set $L_1^*=L_2^*=L$ and compute the first estimate of $\sigma_\mathrm{2,crit}$ from (\ref{mj22}), rewritten for the square geometry as
\begin{equation}\label{mj22x}
    \sigma_\mathrm{2,crit}^{(\Delta p)} = \frac{(tL_1^*L_2^*\,\Delta p_\mathrm{crit})^2-144k^2}{12ktL\left(L/5+L_2^*\right)}
\end{equation}
\noindent Subsequently, we evaluate the changes of chord lengths from Equations (\ref{app24x})--(\ref{app25x}), update the chord lengths $L_1^*=L_1+\Delta L_1^*$ and $L_2^*=L_2+\Delta L_2^*$ and use these in Equation (\ref{mj22x}) to get an improved estimate. This is repeated until the required accuracy is attained. The results of this process match the beam model numerical results very closely, as shown by the correlation of the red square marks in Figure \ref{fig:psig} to the blue triangle marks.


\subsubsection{Critical state curve of the internal instability and a comparison to a plane strain model}
\label{sec:csl}

The critical state formula can be interpreted as a mathematical description of a critical state curve in the $(\Delta p, \sigma_2)$ space. More specifically, the dimensionless version given by Equation (\ref{eq:mj23}) leads to a universal normalized quadratic function independent of cell geometry and material parameters. It is, however, expected that different unit cell geometries might result in different accuracy of the analytical solution, which relies on several key assumptions concerning the beam slenderness and the negligibility of joint dimensions.

The accuracy of the analytical model can be quantified by comparing numerical solutions obtained for different geometries from a detailed two-dimensional plane-strain finite element simulation (for detailed methodology, we refer the reader ahead to Section \ref{sec:model}). This simulation, in contrast to the previous verifications performed on a numerical beam model, abandons the assumptions of beam theory altogether. The numerical simulations were performed on a single periodic unit cell, using a fixed beam length of $L=\SI{6}{\meter}$ and several values of beam height $h$. To account for the plane strain conditions in comparison with the beam theory, the plane strain modulus of $E/(1-\nu^2)$, with $\nu$ the Poisson's ratio, was used in the plane strain simulation.

\begin{figure}[t]
    \centering
    \includegraphics[width=\linewidth]{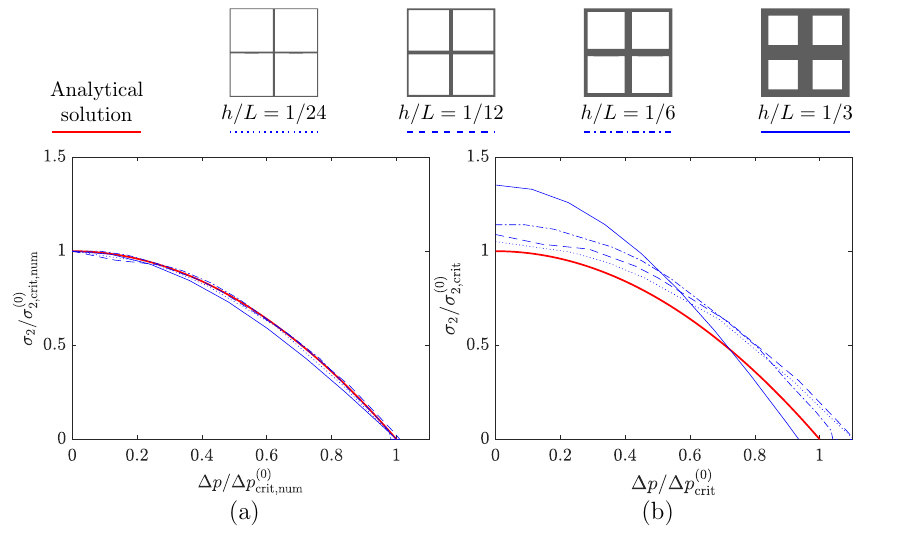}
    \caption{Dimensionless critical state curves for 2D plane-strain simulations of different cell geometries, characterized by the in-plane beam thickness $h$ and its ratio to beam length $L$. (a) Macroscopic stress and pressure difference normalized by their critical values, $\sigma_\mathrm{2,crit,num}^{(0)}$ and $\Delta p_\mathrm{crit,num}^{(0)}$, determined for each case numerically. {The correlation of the curves demonstrates qualitative agreement in the quadratic nature of the critical state curves between the analytical model (in red) and numerical simulations (in blue), regardless of cell geometry}. (b) Macroscopic stress and pressure difference normalized by the analytical predictions of their critical values. {The analytical model (in red) underestimates the critical values of pneumatic pressure and macroscopic stress compared to the numerical simulations (in blue), with the error growing for cell geometries with thicker ligaments. Among sources of this error are beam theory assumptions, neglect of shear effects in the analytical model, and an assumption of infinitesimal joint size when analytically evaluating the pneumatic load.}}
    \label{fig:csl}
\end{figure}

The general shape of the critical state line is predicted by the analytical model very well, as revealed by comparing the predicted thick red curve to the numerical results in blue in Figure \ref{fig:csl}{a}, where the critical state curves are all normalized by their respective critical values of pressure difference (at zero stress) and macroscopic compressive stress (at zero pressure difference). In Figure \ref{fig:csl}{b}, the curves are normalized by the analytical predictions of $\Delta p_\mathrm{crit}^{(0)}$ and $\sigma_\mathrm{2,crit}^{(0)}$ evaluated from Equations~(\ref{mj26}) and~(\ref{mj27}). It turns out that these simple estimates, in general, underestimate the critical pressure difference and macroscopic stress values by up to 30~\% for stocky geometries with a larger $h/L$ ratio. {Furthermore, in the case of the stocky geometries, the error is larger in the left side of the diagram, i.e., for the cases with larger macroscopic stress loading and lower pneumatic pressure magnitude.}

There are several sources of inaccuracy in the analytical model, some competing with each other and some differently pronounced depending on whether pneumatic loading or macroscopic compression is dominant. The behavior of slender geometries ($h/L=1/24$, $h/L=1/12${, dotted and dashed blue curves, respectively}) tends to be captured more accurately, as expected. For stockier beams the inaccuracy can be attributed to several sources. Neglecting shear effects in the beam formulation is consequential, as is equally the assumption of full theoretical length between joints. The latter causes several inaccuracies; firstly, the pneumatic pressure loading is overestimated, as it is considered to be applied to a longer part of the beam than in reality, and secondly, the bulky joints tend to behave partially like rigid bodies connected by shorter beams rather than like joints between two bending beams. Particularly this latter effect dominates for the very thick beams ($h/L=1/6$, $h/L=1/3${, dash-dotted and solid blue curves, respectively}). To corroborate this explanation, we performed additional numerical simulations (results not shown here) using the nonlinear finite element beam model developed in \cite{JLMRH21nonlinbeam}, which we extended for this purpose to allow rigidly fixing parts of the beam length. We have found that, indeed, by fixing the true size of the joint to be rigid or, alternatively, allowing the full theoretical beam length to deform, one might construct upper and lower bounds on the results of the plane-strain simulations presented in Figure~\ref{fig:csl}. 

In conclusion, the analytical model provides an accurate qualitative prediction of the unit cell behavior, with additionally reasonable quantitative accuracy achieved for slender enough unit cell geometries. The loss of quantitative accuracy for stocky geometries was expected given the assumptions made in the construction of the model. Therefore, for further examination of stockier geometries, as well as larger unit cells, we are going to use a plane-strain numerical model instead, using analytical considerations only for the qualitative explanation of the unit cell behavior.

\section{Geometrical and numerical model of a pneumatically actuated metamaterial}
\label{sec:model}

The internal buckling behavior described by the analytical model developed in Section \ref{sec:analytics} is associated with a significant loss of effective stiffness of the microstructure under macroscopic compression, as is also common for patterning behavior of similar microstructures with circular voids \citep{Bertoldi2008Boycemat}. Unlike this classical example, the induction of the pattern in the rectangular void microstructure by pneumatic pressure enables selecting the periodicity of the pattern and, thus, the magnitude of the stiffness loss. The analytical model, as presented so far, was limited to a unit cell with $2\times2$ identical rectangular voids; its findings, however, can be qualitatively if not quantitatively extrapolated also to larger periodic cells. To this goal, consider a unit cell of $2\times4$ rectangular voids, as shown in Figure \ref{fig:motivation_RVE_and_patterns}. In this case, it is possible to apply pressurizing schemes in which positively and negatively pressurized voids alternate with varying periodicity (Figure \ref{fig:motivation_RVE_and_patterns} shows two cases of the periodicity parameter $\eta \in \{1,2\}$). Engineering intuition suggests the stiffness of the buckled microstructure to depend on the magnitude of the horizontal deflection of the vertical column beams. Together with the fully elastic and reversible nature of the patterning process, this allows for the construction of an active metamaterial with switchable stiffness, where a single microstructure can be subjected to different pressurizing schemes, effectively preselecting the post-buckling stiffness.

\begin{figure}[t]
    \centering
    \includegraphics[width=.6\linewidth]{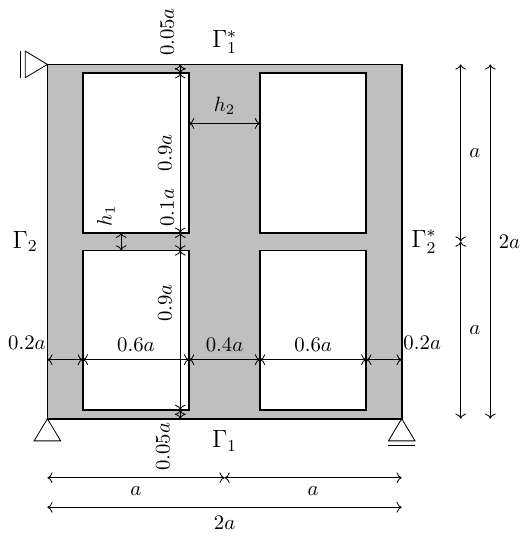}
    \caption{Geometry of a $2\times2$ unit cell for the $\eta=1$ periodicity of pressure actuation. The rectangular shape of voids ($0.6a \times 0.9a$) in a square lattice ($a\times a$) results in vertical column beams thicker than the horizontal ligament beams ($h_2 = 0.4a > h_1 = 0.1a$). Periodic boundary conditions are enforced, with the boundary $\Gamma_1^*$ being a periodic image of the boundary $\Gamma_1$ and likewise the boundary $\Gamma_2^*$ being a periodic image of the boundary $\Gamma_2$. Additionally, pointwise Dirichlet boundary conditions are employed to prevent rigid body motions and the lateral movement of the top left corner.}
    \label{fig:unitcellgeometry}
\end{figure}

We examine this concept by numerical simulations (discussed in Section~\ref{sec:simulations}) of the aforementioned microstructure with rectangular voids. The geometrical model of a typical cell is pictured in Figure \ref{fig:unitcellgeometry}. The rectangular voids are arranged into a square lattice with side $a$. The rectangles are oriented vertically such that the vertical column beams have depth $h_2 = 0.4a$ and the horizontal ligament beams have depth $h_1 = 0.1a$. The $2\times2$ typical cell pictured in Figure~\ref{fig:unitcellgeometry} can be considered a representative volume element (RVE) suitable for examining the pressurizing scheme with the periodicity parameter $\eta = 1$. The pressurizing schemes with larger periodicity parameter require larger RVEs obtained by stacking the same cell vertically. This leads to seemingly unnecessary horizontal ligaments, which are not loaded since the same pressure difference is applied on both sides. Their presence, however, cannot be avoided if the goal is a microstructure to which different pressurizing schemes may be applied. In all cases, the unit cell is considered a microstructural RVE with periodic boundary conditions applied to all sides. Apart from those, rigid body motions are prevented by fixing the lower left and lower right corners, and additionally macroscopic shear deformation is prevented by a horizontal boundary condition on the upper left corner. This simulates lateral fixing of the (in practical application finite and much larger than the presented RVE) structure to prevent macroscopic vertical compression causing a global shear deformation irrespective of the induced pattern.

Focusing on the effective vertical stiffness, we selected a geometry with thicker vertical columns ($h_2 > h_1$). In the pre-buckling phase, the vertical macroscopic stiffness is mostly the product of the axial stiffness of the columns. In the post-buckling phase, it is beneficial for the columns to deform laterally as much as possible in order to achieve a larger stiffness decrease. This is aided by the low axial stiffness of the horizontal ligaments. These ligaments, however, need to retain at least some bending stiffness to be able to resist the applied pneumatic load.

The patterning metamaterial is considered to be a soft polymer slab perforated by rectangular voids. We model the material of the slab using a compressible neo-Hooekan hyperelastic material law in the form presented by
\cite{rivlin1948neohookean}. The strain energy density $W_{\mathrm{NH}}$ defining the compressible neo-Hookean model is given as
\begin{eqnarray}
\label{eq:neohookeanenergydensity}
    W_{\mathrm{NH}} (\bm{F}) &=& \frac{\mu}{2} \left(I_1 (\bm{F})  - 3 -2\log{J(\bm{F})}\right) + \frac{\lambda}{2}\left( J(\bm{F}) - 1\right)^{2}
\end{eqnarray}

\noindent where $\bm{F}$ is the deformation gradient, $I_1 = \mathrm{tr}(\bm{F}^{\mathrm{T}}\bm{F})$ is the first invariant of the right Cauchy-Green deformation tensor $\bm{C} =\bm{F}^\mathrm{T}\bm{F}$, and $J = \det{\bm{F}}$ is the determinant of the deformation gradient. The material parameters $\lambda$ and $\mu$ correspond in the small strain limit to the standard Lamé coefficients. In our study, we use the Young's modulus $E$ and the Poisson's ratio $\nu$ as the primary parameters; they are linked to the Lamé coefficients by

\begin{equation}
    \lambda = \frac{E\nu}{(1-2\nu)(1+\nu)} \quad\quad \mu = \frac{E}{2(1+\nu)} 
\end{equation}

Since all stress- or stiffness-like quantities are normalized by the Young's modulus $E$, its choice is arbitrary. On the other hand, the Poisson's ratio is set to $\nu = 0.499$ in all simulations, which is a typical value for a rubber-like, elastomer material \citep{Mott2009Poissonrationvalues}. Despite this leading to a nearly incompressible hyperelastic material, we do not find any numerical issues with the formulation.

To apply the pressure differences $\Delta p$ and to deal with the self-contact of the structure, we use a third medium method recently developed in \cite{Faltus20243M}, in which the voids are meshed and the void space is represented as a fictitious third medium described by a specific hyperelastic material model. This material model is also based on a neo-Hookean formulation, albeit with very compliant response, enriched by additional second-gradient terms that regularize its behavior and by pneumatic terms introducing a prescribed hydrostatic Cauchy stress. As a result, this material model enforces a prescribed pneumatic pressure difference in the voids while taking care of the internal contact upon void closure as a result of the stiffening of the third medium upon volumetric compression.

Computations are performed using the finite element method (FEM) on a plane-strain RVE model. The choice of the plane strain assumption is motivated by the conditions in the middle layer of a thick polymer slab; it also corresponds to a possible experimental setup in which the metamaterial would be fixed between a pair of PMMA plates to facilitate pneumatic actuation of the voids~\citep{Faltus20243M}. Both the bulk and void regions are discretized by triangular finite elements with quadratic shape functions. To solve for equilibrium of the discretized problem, a modification of the Newton-Raphson solver is used {based on modified Cholesky decomposition. T}he LDLT decomposition of the global stiffness matrix is computed in each Newton iteration and negative entries of the resulting diagonal matrix $\mathsf{D}$ are multiplied by $-1$. {Presence of these negative diagonal entries is correlated with negative eigenvalues of the original matrix, i.e., with instability of the system. The flipping of their sign} prevents the Newton-Raphson method from converging to local maxima and saddle points, thereby converging to locally stable configurations even in the vicinity of bifurcation points, {albeit} at the cost of an increased number of iterations~\citep{Cheng1998modifiedcholesky, Nocedal2006numericaloptimization}. All simulations are run in the MATLAB~\citep{MATLAB2023} environment, with some parts of the code written in C++ to enhance its performance. The utilized code is an extension to an in-house codebase used by \cite{Rokos2020newtonsolver} and \cite{Rokos2020honeycombs}.

\section{Simulation results}
\label{sec:simulations}

\subsection{Patterns for different pressurizing schemes}

\begin{figure}[t]
    \centering
    \begin{tabular}{c c}
         \includegraphics[height = 0.232\textheight]{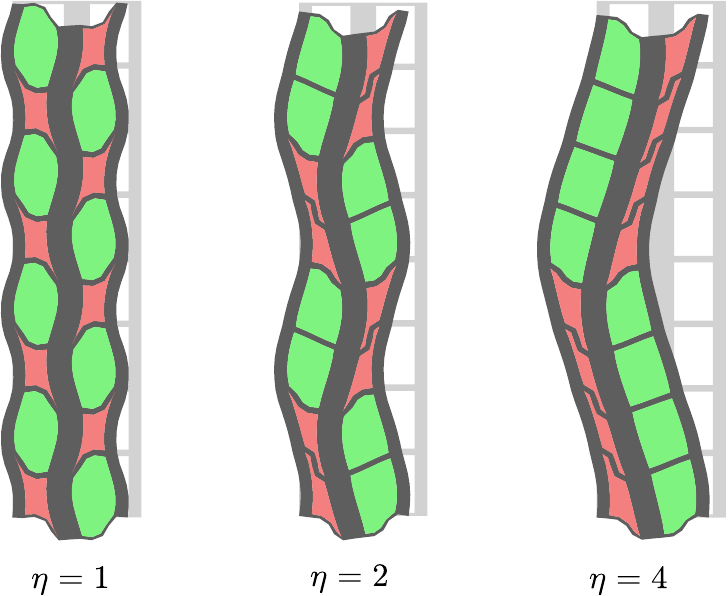}
         &
         \includegraphics[height = 0.232\textheight]{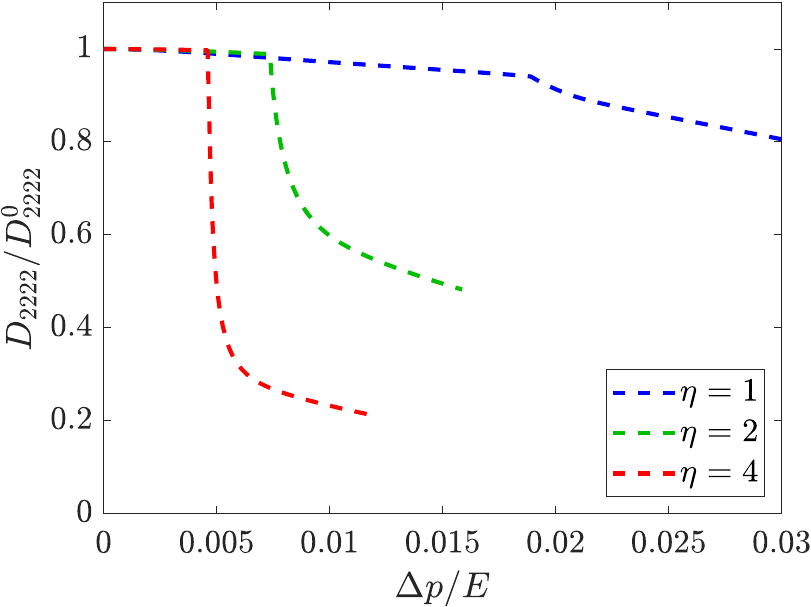}
         \\
         \footnotesize{(a)} & \footnotesize{(b)}
    \end{tabular}
    \caption{Triggering internal patterns in a pneumatically actuated square lattice with rectangular compartments:  (a) Pressurization schemes in the $2\times8$ periodic RVE and deformed shapes of the microstructure at the onset of patterning for each pressurization scheme $\eta$. (b) Dependence of effective macroscopic stiffness component $D_{2222}$ (normalized by the initial stiffness $D^0_{2222}$ of the reference configuration) on the introduced pressure difference $\Delta p$ (normalized by the bulk material Young modulus $E$). Note that $\Delta p$ represents the pressure difference introduced into positive compartments (green compartments in (a)); it is understood that simultaneously $-\Delta p$ is being introduced into negative compartments (red compartments in (a)).}
    \label{fig:patterns_and_their_stiffness}
\end{figure}

Pneumatic loading of the microstructure with the rectangular geometry pictured in Figure \ref{fig:unitcellgeometry} results in a loss of stability and internal patterning similar to that described analytically for a square geometry in Section~\ref{sec:analytics}. With periodic boundary conditions enforced on the whole RVE, the periodicity of the developed pattern is given by the periodicity of the pressurizing scheme used. Figure \ref{fig:patterns_and_their_stiffness} shows three different pressurizing schemes with periodicity parameters $\eta = 1,2,4$ and the deformed state of a $2\times8$ RVE upon pressurization according to these schemes. For the introduction of air pressure, it is assumed that the green compartments in Figure~\ref{fig:patterns_and_their_stiffness}a are pressurized to a pressure difference of $\Delta p > 0$ (inflation) simultaneously with the red compartments being depressurized to a pressure difference of $-\Delta p$ (suction). No other loading is considered during pressurization. The three different pressurizing schemes result in distinct developments of the macroscopic stiffness of the RVE with increasing pneumatic pressure, see Figure \ref{fig:patterns_and_their_stiffness}b. The value of the macroscopic tangent stiffness has been obtained at each time step of the finite element simulation via a computational homogenization scheme \citep{kouznetsova2001approach}.

\begin{table}[h]
    \centering
    \begin{tabular}{c c c}
        \hline
         $\eta$ & $\Delta p_\mathrm{crit}/E$ & $D^\mathrm{post}_{2222}/D^0_{2222}$
         \\ \hline
         1 & $0.016$ & $\sim 90 \%$ 
         \\ 
         2 & $0.006$ & $\sim 65 \%$ 
         \\ 
         4 & $0.004$ & $\sim 30 \%$ 
    \end{tabular}
    \caption{Critical pressure differences $\Delta p_\mathrm{crit}$, expressed relative to $E$, leading to the internal patterning instability and the associated reductions in the vertical component of the effective macroscopic stiffness tensor $D^\mathrm{post}_{2222}$, relative to the reference state of the microstructure $D^0_{2222}$, for three periodicity parameters $\eta$ shown in Figure \ref{fig:patterns_and_their_stiffness}.}
    \label{tab:criticalpressures}
\end{table}

Depending on the pressurization scheme used, both the critical pressure difference $\Delta p_\mathrm{crit}$ necessary to buckle the microstructure and the magnitude of the stiffness loss vary, with larger $\eta$ leading to lower critical pressure differences and larger losses of stiffness; see Table \ref{tab:criticalpressures}.

The resulting behavior can be explained by the shape of the deformation patterns. In general, the pattern can be described as a chord rotation of both columns and horizontal ligaments, the subsequent deformation thus allowing the rectangular voids to either close or rotate based on the sign of their pressurization, consistently with the description of the buckling mechanism presented above in Section \ref{sec:physical_meaning}. As a result, the vertical columns locally bend and form a periodic zig-zag shape, the period of which correlates to the period of the introduced pressurization scheme. Larger periods lead to larger horizontal deflections of the vertical beams, causing larger compliance under vertical compression.

To summarize, pressurizing a square lattice microstructure with thick columns and thin connecting horizontal ligaments opens up a way of inducing an on-demand buckling pattern, the precise shape of which can be controlled by the pattern of the pressure actuation. This leads to a drop in vertical macroscopic stiffness of the material down to 30~\% of the reference value. It can be reasonably expected that using a larger RVE with a larger period of the pressurization scheme could lead to even larger reductions.

\subsection{Stiffness range in loading space}

As indicated already by the analytical model in Section \ref{sec:pressure_and_vertical_stress}, the patterning instability arises from an interplay between the two loading modes: pneumatic pressure and vertical macroscopic loading. This macroscopic loading can be applied in finite element simulation as well, represented by a vertical force load on the top left control node of the cell. Due to the periodic boundary conditions, a force load in this control node represents equally the effective macroscopic first Piola-Kirchhoff stress~$\bm{P}$~\citep{kouznetsova2001approach}. Specifically in our case, the vertical normal component $P_{22}$ is imposed.

\begin{figure}
    \centering
    \includegraphics[width=\linewidth]{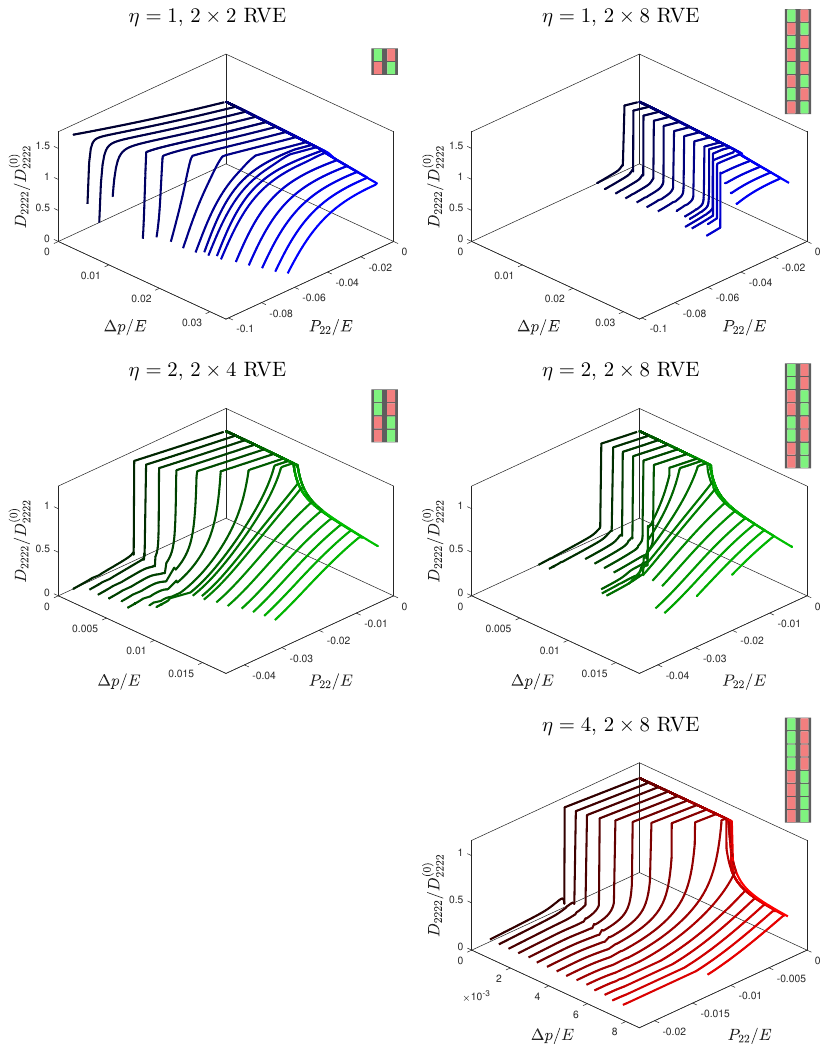}
    \caption{Evolution of component $D_{2222}$ of the macroscopic effective stiffness tensor as a function of prescribed pressure difference $\Delta p$ and macroscopic vertical load $P_{22}$ for the three pressurization schemes $\eta = 1, 2$ and $4$. For each pressurization scheme, two graphs are shown, one of them calculated on the smallest possible RVE for each scheme, i.e., a $2\times2\eta$ unit cell, (left), and the other on a $2\times8$ RVE same for all schemes (right).
    }
    \label{fig:surfaces}
\end{figure}

\begin{figure}
    \centering
    \includegraphics[width=.75\linewidth]{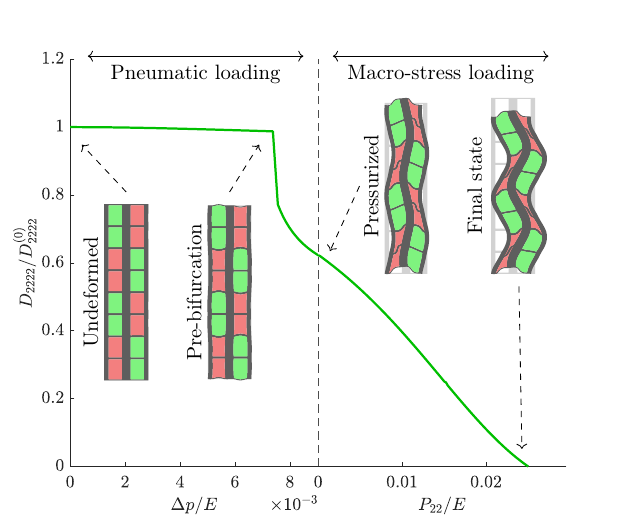}
    \caption{Typical loading path for a $2\times8$ RVE with $\eta = 2$. First, pneumatic pressure difference is introduced up to $\Delta p/E = 0.9 \times 10^{-3}$ at the limit of the first horizonal axis. Since this value exceeds the critical pressure difference, the microstructure buckles into a pattern dictated by the pressurization scheme. During subsequent loading of the already buckled microstructure by vertical macroscopic stress $P_{22}$, this pattern is maintained as the macroscopic vertical stiffness gradually decreases. In this case, the simulation is ended when macroscopic stiffness reaches zero.}
    \label{fig:path}
\end{figure}

Applying the combined pneumatic and macroscopic stress loading to the numerical metamaterial simulations, it is possible to construct graphs of macroscopic stiffness in the ($\Delta p$, $P_{22}$) space. Such graphs are pictured in Figure \ref{fig:surfaces} for the pressurizing schemes $\eta = 1$, $2$, and $4$. To construct these graphs, a number of simulations were performed, in which the given microstructure was first pressurized to a given pneumatic pressure difference level and then loaded by vertical stress load, observing the value of the vertical component of the macroscopic stiffness throughout; Figure~\ref{fig:path} depicts a typical loading path for $\eta=2$ in more detail with the evolution of the microstructure's deformed state.

\begin{figure}
    \centering
    \includegraphics[width=0.6\textwidth]{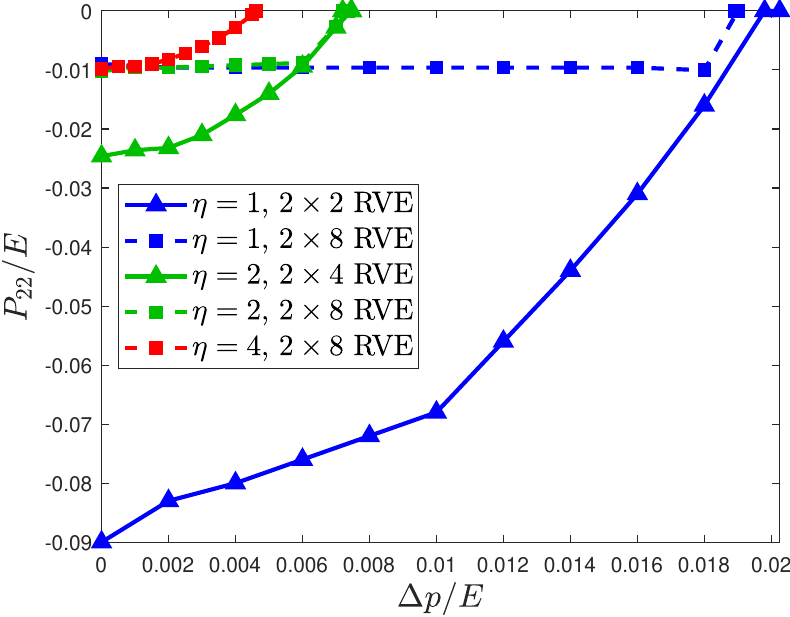}
    \caption{Numerically computed critical state curves of the patterning instability on the proposed metamaterial geometry plotted in the ($\Delta p$, $P_{22}$) space. For each pressurizing scheme $\eta = 1$, $2$, and $4$, the curves have been computed both on an RVE of $2\times2\eta$ cells (the smallest possible for each given scheme) and on an RVE of $2\times8$ cells. In the latter case, the loss of stability associated with macroscopic stress loading occurs at a larger wavelength and thus for a smaller critical macroscopic stress; this is reflected in the apparent cutoff of the stable area within the critical state curve in comparison to the former case.}
    \label{fig:cslsforRVEs}
\end{figure}

Note that the graphs in Figure \ref{fig:surfaces} are smooth and valid for loading of the elastic material in any direction on the ($\Delta p$, $P_{22}$) plane. However, the emergence of an internal pattern is accompanied by a jump from a relatively flat region of stiffnesses close to the reference value to another region with a significantly reduced stiffness. The boundary at which this occurs is in fact the critical state curve as described in Section \ref{sec:csl} (compare Figure~\ref{fig:csl} and Figure~\ref{fig:cslsforRVEs}, which shows the critical state curves for the simulations discussed here).

For each pressurizing scheme, two variants of the graph in Figure \ref{fig:surfaces} illustrate two different situations as to the size of the RVE used: either the smallest possible RVE is used, i.e., $2\times2\eta$ compartments for each pressurizing parameter $\eta$, or a $2\times8$ compartments RVE is used for every scheme. The latter case results in a larger wavelength of the buckled shape in response to macroscopic stress; this in turn leads to a cutoff of the critical state curve of the internal instability and thus a reduction of the stable area bounded by the critical state curve; this is illustrated more closely in Figure~\ref{fig:cslsforRVEs}. This behavior has not been captured by the analytical model presented above in Section \ref{sec:analytics} due to its limitations with respect to the cell size.

This tendency of the microstructure to select a buckling shape with the largest possible wavelength in response to macroscopic stress loading has potentially large implications for the ultimate goal of variable stiffness reduction directed by the applied pressurizing scheme. The introduced pneumatic pressure difference has to be high enough to effectively prevent this from happening. Viable loading paths are especially those where the desired pattern is triggered entirely by the pneumatic loading before the introduction of any macroscopic stress, such as in the simulation presented in Figure \ref{fig:path}. This most commonly expected actuation procedure leaves a moderately large design space in which the pattern is maintained and the stiffness remains low. The interplay between pneumatic and macroscopic loading further paves the way for finer tuning of the exact stiffness value with adjustments to the pneumatic load even during the macroscopic loading.
{ }


\section{Conclusions}
\label{sec:conclusions}

In this work we have investigated pneumatic actuation of a pattern-forming metamaterial microstructure with rectangular voids. Our analytical model based on beam theory demonstrates that patterned pneumatic actuation leads to the emergence of a local pattern otherwise unobserved under classic uniform macroscopic loads. An extension to this model reveals that if the pneumatic loading governed by the pressure difference $\Delta p$ is combined with macroscopic stress $\sigma_2$, the patterning instability is caused by an interplay between the two loading modes, forming a continuous critical state curve in the $(\Delta p, \sigma_2)$ space.

It has further been revealed through numerical simulations that this behavior can be exploited to construct a metamaterial with active control of stiffness response to macroscopic loading. The choice of pressurization scheme effectively dictates the shape of the internal pattern and, thus, also the magnitude of the stiffness loss associated with the patterning. The behavior of the material as to the patterning instability is also fully reversible and independent of the loading path, which reveals further options to control the material behavior through pressurization and depressurization of the voids.

Further development of this concept could lead to more complicated microstructural geometries. At present, we limited ourselves to rectangular voids, for which the macroscopic stiffness of the metamaterial is dictated by the thick columns. A further increase of the stiffness range is limited by the presence of the horizontal ligaments connecting those columns. The design of these ligaments is a compromise between two conflicting requirements: they should be as axially compliant as possible in order not to impede the lateral deformation of the columns, yet they should also possess a significant stiffness to withstand the pressure loading in the adjacent compartments. In the future, it might be possible to overcome this contradiction by a more complicated design of these ligaments, obtained, e.g., by topology optimization. We took the first steps in this direction by unifying pneumatic actuation and contact in a third medium model~\citep{Faltus20243M} which is a convenient formulation for topology optimization of pneumatically actuated structures \citep{Caasenbrood2020pneumatictopology, Dalklint2024inflatablesoftrobots}.

On a similar note, it can be reasoned that the design principle of the bending columns might be achievable by a different actuation method, without the use of patterning instability. This would likely remove the jumping behavior between two different stiffness levels in favor of a smoother, finely tunable slope. Among possibilities how to achieve this we envision (i) actuating the horizontal ligaments themselves rather than the voids, either pneumatically or, e.g., magnetically, or (ii) introducing a different kind of mechanical actuator. An actuator introducing a force oriented diagonally across the square voids should be capable of pushing the microstructure into a deformed shape similar to that of the internal pattern presented here.

The presented results have been focused so far on two-dimensional finite element simulations of a periodic microstructure. Even though the model is two-dimensional, experimental verification should be possible; see, for instance, the experimental setup in \cite{Faltus20243M}, in which a similar experiment was performed on a simple geometry with circular voids. Additionally, despite pattern-forming metamaterials {being} usually presented as 2D microstructures, {three-dimensional desigs are also known in literature~\citep{Shim2012buckliball3Dpattern,Babaee2013bertoldiauxetic3Dpattern,Li2021rudykhsoftinstabreview}. The present} concept could {also} be {extended} to three dimensions. {For instance, a hexahedral lattice with enclosed cavities would avoid some practical limitations of the 2D design}, in particular those related to preventing out-of-plane deformation and ensuring airtight out-of-plane fixation without leakage of pneumatic pressure. {Alternatively, a rotational extrusion of the presented 2D geometry is conceivable, creating a cyllindrical 3D cell}. {T}he formulation of pressurization schemes {on these 3D microstructures} would likely become more difficult, {however, and boundary effects might play a decisive role on the patterning behavior in the case of finite 3D samples.}

\paragraph{Acknowledgement}
 This work has been supported by the Czech Science Foundation [grant no.~GA19-26143X]. The authors would like to thank Chris Verhoeven, a master student at the Eindhoven University of Technology, for his contributions to the FE code development.

\appendix
\setcounter{figure}{0}
\setcounter{table}{0}   
\section{Derivation of the analytical  model of a pressurized unit cell based on beam theory}
\label{app:unit_cell_analytical_model}

The analytical model used for predicting the critical pressure difference is based on the replacement of the solid part of the microstructure by axially incompressible beams. For stability analysis, it is necessary to develop an expression for the elastic energy stored in a typical beam and for the change of length measured along the chord. In a corotational coordinate system, in which the $x$ axis is aligned with the chord, the deformed state of the beam is characterized by centerline deflections $w(x)$ and longitudinal displacements $u(x)$ relative to the chord. The assumption of axial incompressibility/inextensibility leads to the constraint
\begin{equation}\label{app1}
  (1+u'(x))^2 + w'^2(x) = 1  
\end{equation}
 where the prime indicates the derivative with respect to the $x$ coordinate along the chord. For moderate deformation, $u'^2$ can be neglected in comparison with $u'$, but $w'^2$ is considered to be of the same order of magnitude, so that the approximate form of constraint (\ref{app1}) yields
\begin{equation}
  u'(x) = -\frac{1}{2} w'^2(x) 
\end{equation}
 Integration along the chord results in
\begin{equation}\label{app3}
 u(L)-u(0) = \int_0^L u'(x)\,{\rm d}x = -\frac{1}{2}\int_0^L w'^2(x) \,{\rm d}x
\end{equation}
which can be interpreted as the change of the chord length. 

Since the beam is considered as axially incompressible and the shear
distortion is neglected as well, the potential energy stored by elastic deformation comes only from bending and is given by
\begin{equation}\label{app4}
 \mathcal{E}_\mathrm{int} = \frac{EI}{2}\int_0^L w''^2(x)\,{\rm d}x 
\end{equation}
 where $EI$ is the flexural stiffness of the beam cross-section.

For the present purpose, the deflection function $w(x)$ will be approximated by a linear combination of three predefined functions:
\begin{eqnarray}\label{app5}
    N_a(x) &=& -\frac{x^3}{L^2}+\frac{2x^2}{L}-x \\
    N_b(x) &=& -\frac{x^3}{L^2}+\frac{x^2}{L} \\
    N_p(x) &=& \frac{x^4}{L^3}-\frac{2x^3}{L^2}+\frac{x^2}{L} 
\end{eqnarray}
 The first two cubic functions are solutions to ODEs pertinent to the Bernoulli-Navier beam theory with unit rotation prescribed at one beam end and zero rotation prescribed at the other, as well as zero deflection prescribed at either end. The quartic function $N_p$ describes the deformed shape corresponding to a uniformly distributed load perpendicular to the chord in the undeformed state, as well as clamped boundary conditions on both ends. The corresponding deflection approximation
\begin{equation}\label{app8}
w(x) = \theta_a N_a(x) + \theta_b N_b(x) + \tilde{p}_{ab} N_p(x)
\end{equation}
 uses discrete parameters $\theta_a$ and $\theta_b$, which represent the rotations of the beam ends with respect to the chord, and $\tilde{p}_{ab}$, which is a dimensionless parameter characterizing the deflection caused by the distributed load. Substituting Equations~(\ref{app5})--(\ref{app8}) into Equations~(\ref{app4}) and (\ref{app3}) and evaluating the integrals, we obtain approximate formulae for the elastic energy
\begin{equation}\label{a9}
    \mathcal{E}^{(ab)}_\mathrm{int}(\theta_a,\theta_b,\tilde{p}_{ab}) = \frac{2EI}{L}\left(\theta_a^2+\theta_a\theta_b+\theta_b^2\right) + \frac{2EI}{5L}\tilde{p}_{ab}^2
\end{equation}
and the change in chord length
\begin{equation}\label{app10}
    \Delta L^*_{ab}(\theta_a,\theta_b,\tilde{p}_{ab}) = -\frac{L}{30}\left(2\theta_a^2-\theta_a\theta_b+2\theta_b^2-\theta_a\tilde{p}_{ab}+\theta_b\tilde{p}_{ab}+\frac{2}{7}\tilde{p}_{ab}^2\right) 
\end{equation}
expressed as quadratic functions of the discrete parameters. Moreover, the volume between the chord and the deformed centerline 
will be approximated by
\begin{equation}\label{a11}
V(\theta_a,\theta_b,\tilde{p}_{ab}) =t\int_0^L w(x)\,{\rm d}x = \frac{tL^2}{12}\left(\theta_b-\theta_a+\frac{2}{5}\tilde{p}_{ab}\right)
\end{equation}
Strictly speaking, this area should be taken as
\begin{equation}
V^* =t\int_0^L w(x)(1+u'(x))\,{\rm d}x = V -\frac{t}{2} \int_0^L w(x)w'^2(x)\,{\rm d}x
\end{equation}
which would lead to a cubic function of the discrete parameters, but as the first approximation we consider the linear function specified in Equation~(\ref{a11}). Note that $V$ is perfectly appropriate if we want to express the energy of uniformly distributed dead loads (taken per unit length of the undeformed centerline), while $V^*$ would be the appropriate work-conjugate quantity for loading by pressure that corresponds to the force per unit deformed area.

Parameter $\tilde{p}_{ab}$ can be considered as a local degree of freedom specified at the beam level which affects the total 
potential energy of the system only by the contribution coming from one beam, given by
\begin{equation}
    \mathcal{E}_\mathrm{p}^{(ab)} = \mathcal{E}_\mathrm{int}^{(ab)}(\theta_a,\theta_b,\tilde{p}_{ab})-\Delta p V(\theta_a,\theta_b,\tilde{p}_{ab})
\end{equation}
where $\Delta p$ is the externally applied pressure difference. 
When the total potential energy is minimized, its partial derivative with respect to $\tilde{p}_{ab}$ is obtained
by differentiating the beam contribution $\mathcal{E}_\mathrm{p}^{(ab)}$,
and the condition of vanishing first derivative leads to a local
equation
\begin{equation}
    \frac{\partial \mathcal{E}^{(ab)}_\mathrm{int}(\theta_a,\theta_b,\tilde{p}_{ab})}{\partial \tilde{p}_{ab}}=\Delta p\, \frac{\partial V(\theta_a,\theta_b,\tilde{p}_{ab})}{\partial \tilde{p}_{ab}}
\end{equation}
Substituting from Equations~(\ref{a9}) and (\ref{a11}), we obtain a linear equation
\begin{equation}\label{a15}
     \frac{4EI}{5L}\tilde{p}_{ab} = \frac{tL^2}{30}\Delta p
\end{equation}
from which
\begin{equation}\label{a15}
     \tilde{p}_{ab} = \frac{tL^3\Delta p}{24EI}
\end{equation}
In this way, the local degree of freedom $\tilde{p}_{ab}$ can be eliminated on the beam level and the contribution to the total potential energy can thus be expressed exclusively as a function of end rotations with respect to the chord and the applied pressure difference:

\begin{eqnarray}
    \mathcal{E}_\mathrm{p}^{(ab)} &=& \mathcal{E}_\mathrm{int}\left(\theta_a,\theta_b,\frac{\Delta ptL^3}{24EI}\right)-\Delta p V\left(\theta_a,\theta_b,\frac{\Delta ptL^3}{24EI}\right)
    \\ \nonumber
    &=& \frac{2EI}{L}\left(\theta_a^2+\theta_a\theta_b+\theta_b^2\right) + \frac{2EI}{5L}\frac{t^2L^6(\Delta p)^2}{(24EI)^2}
    -\frac{tL^2\Delta p }{12}\left(\theta_b-\theta_a+\frac{2}{5}\frac{tL^3\Delta p}{24EI}\right)
\end{eqnarray}
 Since $\Delta p$ plays the role of a prescribed load, terms that depend exclusively on $\Delta p$ do not need to be included in the final expression for the state-dependent part of the potential energy. Therefore, we will consider the contribution of one beam in the form
\begin{equation}
  \mathcal{E}_\mathrm{p}^{(ab)} =    \frac{2EI}{L}\left(\theta_a^2+\theta_a\theta_b+\theta_b^2\right) 
    +\frac{tL^2\Delta p}{12}\left(\theta_a-\theta_b\right)
\label{eq:one_beam_potential_energy}
\end{equation}
 where $t$, $EI$ and $L$ are characteristics of the considered beam and $\Delta p$ is the pressure difference acting on the beam (oriented downwards if subscript $a$ refers to the left end and $b$ to the right end).  

\begin{table}[h]
    \centering
    \begin{tabular}{ccccc}
    \hline
   beam &  $\theta_a$    &  $\theta_b$  &  $\theta_a-\theta_b$ & $\Delta p$\\
   \hline
     1    &  $\varphi-\psi_1$ & $-\varphi-\psi_1$ & $2\varphi$ & $2\Delta p$ \\
          3    &  $-\varphi+\psi_1$ & $\varphi+\psi_1$ & $-2\varphi$ & $-2\Delta p$ \\
               5    &  $-\varphi-\psi_1$ & $\varphi-\psi_1$ & $-2\varphi$ & $-2\Delta p$\\
                    7    &  $\varphi+\psi_1$ & $-\varphi+\psi_1$ & $2\varphi$ & $2\Delta p$ \\
                    \hline
2 & $-\varphi-\psi_2$ & $\varphi-\psi_2$ & $-2\varphi$ & $2\Delta p$ \\
4 & $\varphi-\psi_2$ & $-\varphi-\psi_2$ & $2\varphi$ & $-2\Delta p$\\
6 & $\varphi+\psi_2$ & $-\varphi+\psi_2$ & $2\varphi$ & $-2\Delta p$\\
8 & $-\varphi+\psi_2$ & $\varphi+\psi_2$ & $-2\varphi$ & $2\Delta p$
    \end{tabular}
    \caption{Parameters $\theta_a$, $\theta_b$, and $p$ appearing in the potential energy expression (\ref{eq:one_beam_potential_energy}) for the unit cell model shown in Figure~\ref{fig:analytical_lattice}. Rotations of end sections and pressure loads for horizontal and vertical beams.}
    \label{tab:rotations}
\end{table}

For the individual beams of the periodic cell shown in Figure~\ref{fig:analytical_lattice}, Table~\ref{tab:rotations} summarizes the end rotations with respect to the chord, their differences, and the applied pressures. It turns out that the contribution of each horizontal beam is

\begin{eqnarray}\nonumber
  \mathcal{E}_\mathrm{p}^\mathrm{hor} &=&    \frac{2EI_1}{L_1}\left((\varphi-\psi_1)^2+(\varphi-\psi_1)(-\varphi-\psi_1)+(-\varphi-\psi_1)^2\right) 
    +\frac{2tL_1^2\Delta p}{12}2\varphi \\
    &=& \frac{2EI_1}{L_1}\left(3\psi_1^2+\varphi^2\right) +\frac{ tL_1^2\Delta p}{3}\varphi
\end{eqnarray}

\noindent and the contribution of each vertical beam is

\begin{eqnarray}\nonumber
  \mathcal{E}_\mathrm{p}^\mathrm{ver} &=&    \frac{2EI_2}{L_2}\left((-\varphi-\psi_2)^2+(-\varphi-\psi_2)(\varphi-\psi_2)+(\varphi-\psi_2)^2\right) 
    -\frac{2tL_2^2\Delta p}{12}2\varphi \\
    &=& \frac{2EI_2}{L_2}\left(3\psi_2^2+\varphi^2\right) -\frac{tL_2^2\Delta p}{3}\varphi
\end{eqnarray}
 Note that these expressions contain the effect of the deviation of the deformed centerline from the chord. This is why the additional term reflecting the energy of applied pressure difference is based on volumes evaluated for parallelograms bounded by the chords; those are given by
\begin{equation}
    V_\mathrm{A} = t(L_1+\Delta L_1^*)(L_2+\Delta L_2^*)\cos(\psi_2-\psi_1)
\end{equation}
for the A compartments (under pressure) and
\begin{equation}
    V_\mathrm{B} = t(L_1+\Delta L_1^*)(L_2+\Delta L_2^*)\cos(\psi_2+\psi_1)
\end{equation}
for the B compartments (under suction). Finally, the total potential
energy of the system consisting of the unit cell in Figure~\ref{fig:analytical_lattice} and applied 
alternating pressure difference $\pm\Delta p$ reads
\begin{eqnarray}\nonumber
    \mathcal{E}_\mathrm{p} &=& 4E_\mathrm{p}^\mathrm{ver}+4E_\mathrm{p}^\mathrm{hor} - 2\Delta p\,V_\mathrm{A} +2\Delta p\,V_\mathrm{B} =\\ \nonumber
    &=& \underbrace{\frac{8EI_1}{L_1}\left(3\psi_1^2+\varphi^2\right) + \frac{8EI_2}{L_2}\left(3\psi_2^2+\varphi^2\right)}_{\mathcal{E}_\mathrm{int}} +
    \\
    &+& \underbrace{\frac{4\Delta p\,}{3}t(L_1^2-L_2^2)\varphi - 4\Delta p\,t(L_1+\Delta L_1^*)(L_2+\Delta L_2^*)\sin\psi_1\sin\psi_2}_{\mathcal{E}_\mathrm{ext}}
\end{eqnarray}
 The first two terms represent the energy stored by elastic deformation and they are introduced into the stability analysis via $\mathcal{E}_\mathrm{int}$ in Equation (\ref{mj1}). The last two terms are the energy of the pressure load and they are reproduced as $\mathcal{E}_\mathrm{ext}$ in Equation (\ref{eq:mj2}), where $L_i+\Delta L_i^*$ is denoted as $L_i^*$, $i\in\{1,2\}$.

The increments $\Delta L_1^*$ of the chord length can be evaluated from (\ref{app10}) with $\tilde{p}$ replaced by $tL^3\Delta p/(24EI)$ and $\theta_a$, $\theta_b$ and $\Delta p$ values taken according to the appropriate row in Table~\ref{tab:rotations}. For horizontal beams, $L=L_1$ and $EI=EI_1$, and the resulting change of chord length is
\begin{eqnarray}\nonumber
    \Delta L_1^*&=& -\frac{L_1}{30}\Bigg(2(\varphi-\psi_1)^2-(\varphi-\psi_1)(-\varphi-\psi_1)+2(-\varphi-\psi_1)^2-(\varphi-\psi_1)\frac{2t\Delta p\,L_1^3}{24EI_1} 
    \\ \nonumber 
    &+& (-\varphi-\psi_1)\frac{2t\Delta p\,L_1^3}{24EI_1}+\frac{2}{7}\left(\frac{2t\Delta p\,L_1^3}{24EI_1}\right)^2\Bigg)
    \\
    &=&
    -\frac{L_1}{30}\left(5\varphi^2+3\psi_1^2\right) +\frac{tL_1^4\varphi\,\Delta p}{180EI_1}-\frac{t^2L_1^7(\Delta p)^2}{15120(EI_1)^2}\label{app24}
\end{eqnarray}
 For vertical beams with $L=L_2$ and $EI=EI_2$, we analogously obtain
\begin{equation}\label{app25}
    \Delta L_2^* =  -\frac{L_2}{30}\left(5\varphi^2+3\psi_2^2\right) -\frac{tL_2^4\varphi\,\Delta p}{180EI_2}-\frac{t^2L_2^7(\Delta p)^2}{15120(EI_2)^2}
\end{equation}

\section{Extension of the analytical model to vertical compression}
\label{app:pressure_stress_interaction}

In \ref{app:unit_cell_analytical_model} we have considered only loading by alternating inflation/suction applied in compartments A and B, recall Figure~\ref{fig:analytical_lattice}a. Let us now add the effect of macroscopic normal stress $\sigma_2$ applied to the unit cell in the vertical direction. This stress plays here the role of prescribed external loading, similarly to the applied pneumatic pressure difference $\Delta p$. Therefore, the expression for the energy of external forces needs to be augmented by the term
\begin{equation}\label{eq:appb1}
    \mathcal{E}_\mathrm{ext,\sigma} = -4tL_1L_2\sigma_2 \varepsilon_2 
\end{equation}
 Here, $4tL_1L_2$ is the total undeformed volume of the periodic cell (consisting of four compartments), and
\begin{equation}\label{eq:appb2}
     \varepsilon_2 = \frac{L_2^*}{L_2} \cos\psi_2 -1
\end{equation} 
 is the macroscopic normal strain in the vertical direction, evaluated from the initial cell size, $2L_2$, and the cell size in the deformed state, $2L_2^*\cos\psi_2$, where $L_2^*$ is the chord length in the deformed state and $\psi_2$ is the angle by which the chord
of initially vertical beams deviates from the vertical axis. Recall that $L_2^*$ may be written as $L_2+\Delta L_2^*$ where $\Delta L_2^*$ is a function of $\varphi$ and $\psi_2$ given by Equation~(\ref{app25}). The assumption of beam inextensibility is maintained, and $\sigma_2$ thus does not cause any axial strain in the beams. In Equations~(\ref{eq:appb1}) and (\ref{eq:appb2}) we consider the standard sign convention with positive stress meaning tension and positive strain meaning extension. In Section~\ref{sec:pressure_and_vertical_stress} we investigate the response under compression, with $\sigma_2<0$ and $\varepsilon_2<0$.

For including the applied load $\sigma_2$ in the equilibrium equations we need the first derivatives of $\mathcal{E}_\mathrm{ext,\sigma}$, which are given by 
\begin{eqnarray}\label{appb3}
   \frac{\partial \mathcal{E}_\mathrm{ext,\sigma}}{\partial\varphi} &=&    -4tL_1\sigma_2\frac{\partial L_2^*}{\partial\varphi}\cos\psi_2
   \\
    \frac{\partial \mathcal{E}_\mathrm{ext,\sigma}}{\partial\psi_2} &=&  -4tL_1\sigma_2\left(\frac{\partial L_2^*}{\partial\psi_2}\cos\psi_2-L_2^*\sin\psi_2\right)
\end{eqnarray}
 where the derivatives of $L_2^*$ follow by differentiation of Equation~(\ref{app25}) as

\begin{eqnarray}
    \frac{\partial L_2^*}{\partial\varphi} &=& -\frac{L_2}{3}\varphi -\frac{tL_2^4\,\Delta p}{180EI_2}
    \\ \label{appb6}
  \frac{\partial L_2^*}{\partial\psi_2} &=& -\frac{L_2}{5}\psi_2
\end{eqnarray}
 Note that neither Equation~(\ref{eq:appb1}) nor~(\ref{eq:appb2}) contain any term with $\psi_1$; hence, the corresponding derivative vanishes.

For the stability analysis, we further need the second order derivatives, which are given by
\begin{eqnarray}
   \frac{\partial^2 \mathcal{E}_\mathrm{ext,\sigma}}{\partial\varphi^2} &=&  
   -4tL_1\sigma_2\frac{\partial^2 L_2^*}{\partial\varphi^2}\cos\psi_2
   \\ \label{eq:appb8}
 \frac{\partial^2 \mathcal{E}_\mathrm{ext,\sigma}}{\partial\psi_2\partial\varphi} &=&  -4tL_1\sigma_2\left(\frac{\partial^2 L_2^*}{\partial\psi_2\partial\varphi}\cos\psi_2-\frac{\partial L_2^*}{\partial\varphi}\sin\psi_2\right)
    \\
 \frac{\partial^2 \mathcal{E}_\mathrm{ext,\sigma}}{\partial\psi_2^2} &=&  
  -4tL_1\sigma_2\left(\frac{\partial^2 L_2^*}{\partial\psi_2^2}\cos\psi_2-2\frac{\partial L_2^*}{\partial\psi_2}\sin\psi_2-L_2^*\cos\psi_2\right)
\end{eqnarray}
 Despite the additional mode of loading, the equilibrium equations still admit the fundamental solution characterized by $\psi_1=\psi_2=0$. Thus, in the stability analysis of this solution, $\cos\psi_2$ can be replaced by $1$ and $\sin\psi_2$ by $0$. Moreover, the mixed derivative of $L_2^*$ in Equation~(\ref{eq:appb8}) vanishes and the second order derivatives with respect to $\varphi$ and $\psi_2$ are given by the constants $-L_2/3$ and $-L_2/5$. Therefore, the terms to be added to the stiffness matrix corresponding to the fundamental solution are
\begin{eqnarray}\label{appb10}
 K_{11,\sigma} &=&  \frac{\partial^2 \mathcal{E}_\mathrm{ext,\sigma}}{\partial\varphi^2}\Big|_{\psi_2=0} = 
   \frac{4}{3}tL_1L_2\sigma_2
   \\ \label{appb11}
 K_{33,\sigma} &=& \frac{\partial^2 \mathcal{E}_\mathrm{ext,\sigma}}{\partial\psi_2^2}\Big|_{\psi_2=0} =
  \frac{4}{5}tL_1\left(L_2+5L_2^*\right)\sigma_2
\end{eqnarray}
 These terms are reflected in the matrix presented in Equation (\ref{eq:tangent_pressureandstress}). Similarly to the patterned pressure loading, they have a destabilizing effect for $\sigma_2<0$.

\bibliographystyle{elsarticle-harv}
\bibliography{biblio2.bib}





\end{document}